\documentclass[preprint2,longabstract]{aastex}

\usepackage{graphicx}
\usepackage{color}
\usepackage{longtable}
\usepackage{indentfirst}
\usepackage{txfonts}
\usepackage[flushleft]{threeparttable}

\shorttitle{CO observations towards N10}
\shortauthors{Gama et al.}

\begin{document}

%% LaTeX will automatically break titles if they run longer than
%% one line. However, you may use \\ to force a line break if
%% you desire.

   \title{CO observations and investigation of triggered star formation \\ towards N10 infrared bubble and surroundings}
   %ArXiv v2 

%% Use \author, \affil, and the \and command to format
%% author and affiliation information.
%% Note that \email has replaced the old \authoremail command
%% from AASTeX v4.0. You can use \email to mark an email address
%% anywhere in the paper, not just in the front matter.
%% As in the title, use \\ to force line breaks.

\author{D.R.G. Gama, J.R.D. Lepine and E. Mendoza}
\affil{Departamento de Astronomia do IAG/USP, Sao Paulo - Brazil}

\author{Y. Wu}
\affil{University of Peking, Beijing - China}
\email{ywu@pku.edu.cn}

\and

\author{J. Yuan}
\affil{National Astronomical Observatories of China (NAOC)}

%% Notice that each of these authors has alternate affiliations, which
%% are identified by the \altaffilmark after each name.  Specify alternate
%% affiliation information with \altaffiltext, with one command per each
%% affiliation.

%\altaffiltext{1}{Visiting Astronomer, Cerro Tololo Inter-American Observatory.
%CTIO is operated by AURA, Inc.\ under contract to the National Science
%Foundation.}
%\altaffilmark{1}
%\altaffiltext{2}{Society of Fellows, Harvard University.}
%\altaffiltext{3}{present address: Center for Astrophysics}

%% Mark off your abstract in the ``abstract'' environment. In the manuscript
%% style, abstract will output a Received/Accepted line after the
%% title and affiliation information. No date will appear since the author
%% does not have this information. The dates will be filled in by the
%% editorial office after submission.

\begin{abstract}

  % context heading (optional)
  % {} leave it empty if necessary  
   {We studied the environment of the dust bubble \object{N10} in molecular emission. Infrared bubbles, first detected by the GLIMPSE survey at 8.0 $\mu$m, are ideal regions to investigate the effect of the expansion of the HII region on its surroundings eventual triggered star formation at its borders.}
  % aims heading (mandatory)
   {In this work, we present a multi-wavelength study of \object{N10}. This bubble is especially interesting as infrared studies of the young stellar content suggest a scenario of ongoing star formation, possibly triggered, on the edge of the HII region.}
  % methods heading (mandatory)
   {We carried out observations of $^{12}$CO(1-0) and $^{13}$CO(1-0) emission at PMO 13.7-m towards \object{N10}. We also analyzed the IR and sub-mm emission on this region and compare those different tracers to obtain a detailed view of the interaction between the expanding HII region and the molecular gas. We also estimated the parameters of the denser cold dust condensation and of the ionized gas inside the shell.} 
  % results heading (mandatory)
   {Bright CO emission was detected and two molecular clumps were identified, from which we have derived physical parameters. We also estimate the parameters for the densest cold dust condensation and for the ionized gas inside the shell. The comparison between the dynamical age of this region and the fragmentation time scale favors the ``Radiation-Driven Implosion" mechanism of star formation.}
  % conclusions heading (optional), leave it empty if necessary 
   {\object{N10} reveals to be specially interesting case with gas structures in a narrow frontier between HII region and surrounding molecular material, and with a range of ages of YSOs situated in region indicating triggered star formation.}

\end{abstract}


%________________________________________________________________

\section{Introduction}
\label{Introduction}

%________________________________________________________________

In the last decade the studies about massive star forming regions have gained considerable attention. Questions as whether the interaction of massive stars with their surrounding molecular clouds triggers the star formation have been amply discussed. The discovery of the ``infrared bubbles'', a new type of object first cataloged through the GLIMPSE\footnote{Galactic Legacy Infrared Mid-Plane Survey Extraordinaire} \citep{benjamin2003, churchwell2009} at 8.0 $\mu$m, offers a new powerful tool to investigate the star formation process. Those objects present a bright border at 8.0 $\mu$m, caused by the emission of Polycyclic Aromatic Hydrocarbons (PAHs), excited by ultraviolet radiation (UV), which surrounds a region of ionized gas \citep{churchwell2006,churchwell2007}. 

The bubbles were detected in observations performed by the Spitzer satellite, in a survey that revealed about 600 bright objects at mid-infrared wavelengths \citep{churchwell2006}. Shortly afterwards the Churchwell catalog were complemented by the Milky Way Project (MWP) catalog by \cite{simpson2012}. This is the most recent scientific citizen-generated catalog, where the bubbles were identified by thousands of volunteers and therefore their classification is more reliable. \citet{deharveng2010} have identified and studied a large sample of bubbles and concluded that the shell of each one, detected at 8.0 $\mu$m, is an evidence of an HII region produced by the ionizing massive stars. \citet{thompson2012} suggested that the expansion of the bubbles triggers the formation of Young Stellar Objects (YSOs) which is a non-negligible process in  Galactic scales. \citet{kendrew2012,kendrew2016} also have observed this behavior and, although evidences of triggering via bubble expansion were missing, these authors found populations of YSOs near the borders of expanding bubbles, which could offer us important clues to the star formation process and the expansion of infrared bubbles.

We can interpret these bubbles as basically ionized gas surrounded by cold dust. A Photon-Dominated Region (PDR) in the inner regions of the shell, can be identified and described at mid-infrared wavelengths \citep{lefloch2005}. The PDRs can be the result of the HII region expansion. They can be seen as regions where the ionization front is still progressing in the densest medium of the original cloud, generating an interface between ionized and neutral gases. The larger density in the PDR is possibly due to material collected by the expansion of the HII region. In this region the UV flux decreases sharply, allowing the existence of molecular and grain species. The massive stars interact with the original molecular cloud and, by their UV radiation, generate the interface between ionized gas and neutral gas.

In principle, the structure of the objects, should it be spherical shells or rings, allow us to understand correctly  the kinematics of the gas and the chronology of the newly formed stars. Several works claimed that there was evidence for star formation triggered by the expansion of the HII region \citep{dewangan2012,zavagno2010,beuther2011}. \citet{jinghua2014} found velocity differences of the order of 30 km s$^{-1}$ between distinct parts of the ring in the bubble N6, which would indicate a quite larger expansion velocity than those considered by other authors, of a few km s$^{-1}$ \citep[e.g.][]{beaumont2010}. \cite{dale2005} and \cite{dale2008} carried out simulations to study the effects of stellar feedback in molecular clouds. They suggested that it is not possible to determine if the formation of an YSO was triggered or not by the expansion of the HII region and as a consequence studies of triggered star formation should be done statistically.

Therefore, it is important to gather a number of well studied bubbles to establish if they have similar formation histories, if they have similar morphology and if they give similar answers to the process about triggered star formation. 

In this work we present a detailed study of \object{N10}, a remarkable bubble which has molecular clumps and YSOs associated with its surrounding shell. We present our CO observations and we analyze them together with the relevant data at different wavelengths: 8.0 $\mu$m, PAH emission; 24 $\mu$m, hot grains from ionized region; 870 $\mu$m, dust emission; and 20 cm free-free emission, from hot gas.

This paper is organized as follows: our target object is introduced in Section \ref{N10}. In Section \ref{data} we describe the CO observations and the archival data used in this paper. We dedicate the Section \ref{results} to present our results and the Section \ref{discussion} to discuss these results. Finally, we summarized our mainly conclusions in Section \ref{conclusions}. 

%__________________________________________________________________

\section{The bubble N10}
\label{N10}

%________________________________________________________________

\begin{figure*} [ht]
\begin{center}
% \setcaptionmargin{1cm}
%\includegraphics[scale=0.45,angle=0]{fig01.eps}
\includegraphics[scale=0.47,angle=0]{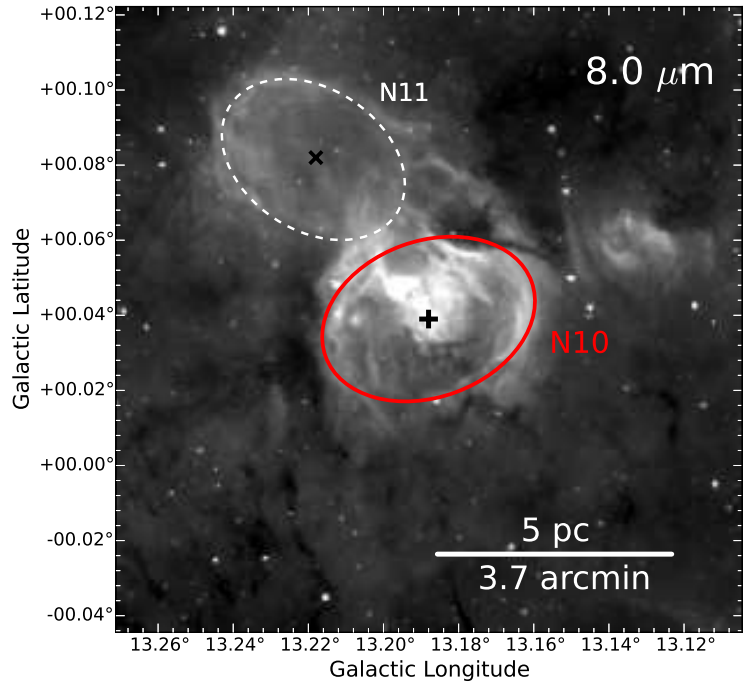}
\caption{Map of the bubble N10, located to a distance of 4.7 kpc, at 8.0 $\mu$m. The black cross indicates the center of the HII region inside the bubble of our interest. The red ellipse indicates the edge of N10. The scale corresponds to a field of 5 pc ($\sim$3.7'). The white dashed ellipse indicates N11 position, which center is located at $l$=13.218$^{\circ}$, $b$=0.082$^{\circ}$, marked by the black ``x''.}
\label{n10_8microns}
\end{center}
\end{figure*}

\object{N10} is situated in the direction $l$ = 13.188$^{\circ}$, $b$ = 0.039$^{\circ}$ \citep{churchwell2006}. This object is also identified as \object{MWP1G013189+000428} by \cite{simpson2012}. It appears in the Spitzer-GLIMPSE 8.0 $\mu$m image as a bright ring-like structure (Figure \ref{n10_8microns}). The 8.0 $\mu$m emission in the border of bubbles is attributed to PDRs containing PAHs; the gas density in the PDR can be much larger (up to a factor 10) than that of the surrounding medium \citep{deharveng2010,churchwell2006}. In Figure \ref{n10_8microns} we show an ellipse to mark the boundaries of the bubble, for later reference in images at different wavelengths. \citet{churchwell2006} consider \object{N10} a bipolar (or double) bubble, since a small bubble (N11) seems to be connected to \object{N10} in the North of it. However, \cite{deharveng2015} recently showed that N10/N11 is not a bipolar bubble, as \cite{churchwell2006} misidentified. In this work we deal only with the case of \object{N10}.

% Fig 01 

The distance of \object{N10} was estimated as 4.9 kpc by \citet{churchwell2006}, 4.1 kpc by \citet{beaumont2010}, 4.6 kpc by \citet{pandian2008} and 4.9 kpc by \citet{watson2008}. These are kinematic distances estimated using different rotation curves, which explain the discrepancies. 

\begin{figure*} [ht]
\begin{center}
\includegraphics[scale=0.5,angle=0]{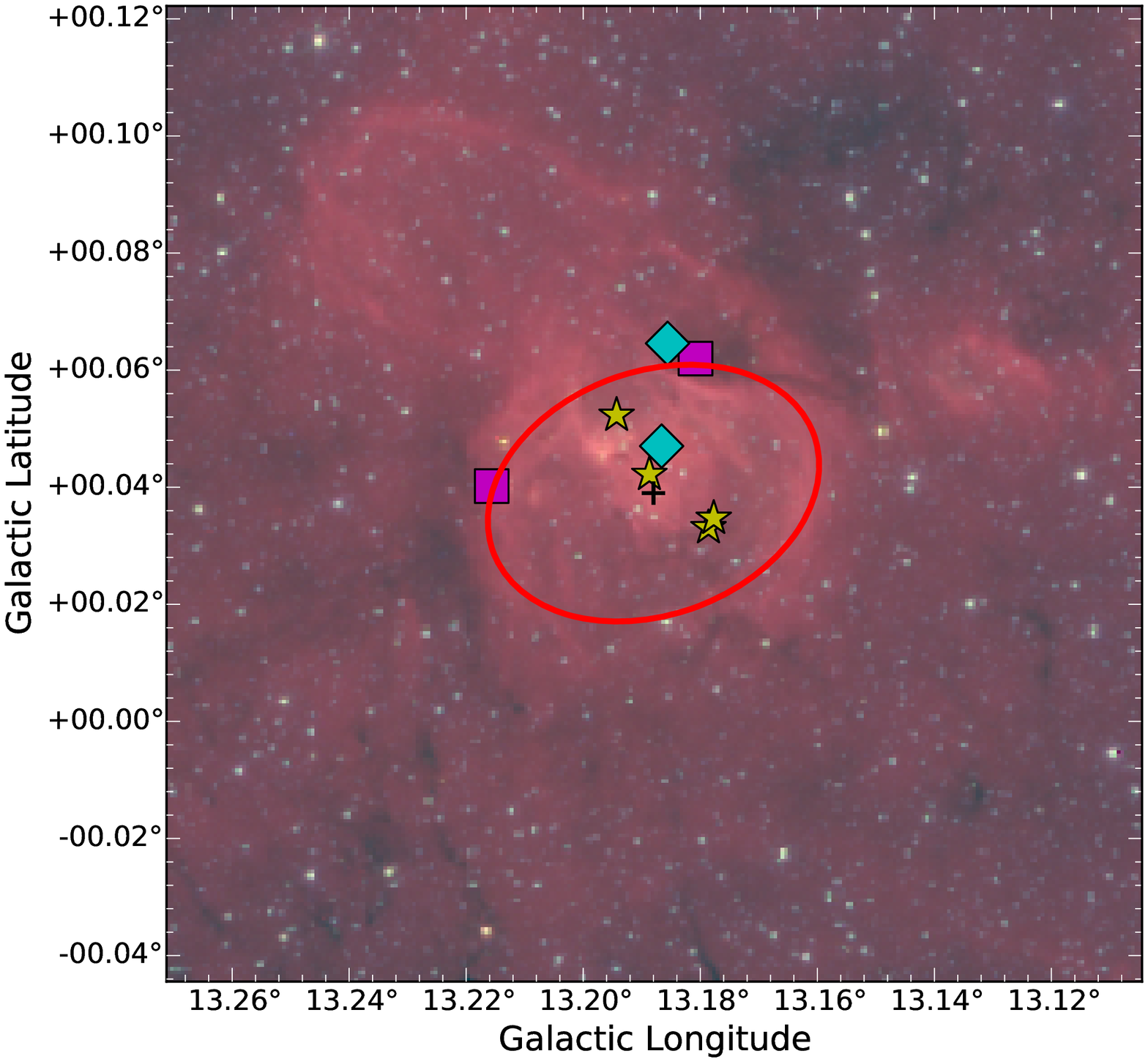}
\caption{Three color composite image of bubble N10 with Spitzer-GLIMPSE 3.6 $\mu$m (blue), 4.5 $\mu$m (green) and 8.0 $\mu$m (red). The positions of the stars from \citet{watson2008} are marked as yellow stars; dust condensations identified by \citet{deharveng2010} are indicated by magenta squares (\#1 at the right side, \#2 at the left side); cyan diamonds represents the position of methanol masers by Szymczak (2000; up) and Pandian (2008; down).}
\label{map8_objects}
\end{center}
\end{figure*}

A methanol (CH$_{3}$OH) maser in \object{N10} region was first reported by \citet{szymczak2000}, detected towards the IRAS 18111-1729 source. It is accepted that the methanol masers (as is the case of the present one) are associated with the earliest stages of massive star formation \citep{minier2002}. Figure \ref{map8_objects} shows this methanol maser source located on the border of one of the two bright 870 $\mu$m condensations adjacent to the bubble. The second CH$_{3}$OH maser reported by \citet{pandian2008} seems to be associated with a SVSS\footnote{NRAO VLA Sky Survey \citep{condon1998}} source.   

In their study of the central region, \citet{watson2008} identified four stars as possible ionizing stars in N10 (see Table \ref{stars}), based on their Spectral Energy Distributions (SEDs) which are well-fitted by a stellar photosphere. The position of stars are also plotted in Figure \ref{map8_objects}. Assuming a radius of 1.61 pc for the densest dark cloud, \cite{ma2013} estimated a dynamical age ${\mathrm t_{dyn} = 9.17 \times 10^{4}}$ yr for \object{N10}. Nevertheless, they argue that this value could be larger since the density of the true ambient where the stars originally were formed could be larger that they considered.

% Table 01
\begin{deluxetable}{c c c c c} 
%\tabletypesize{\scriptsize}
\tablecolumns{5} 
\tablewidth{0pt} 
\tablecaption{Candidates ionizing stars in N10 system.\label{stars}} 
\tablehead{ 
\colhead{ID \tablenotemark{a}} & \colhead{A.R. (J2000)}   & \colhead{DEC (J2000)}    & \colhead{Spectral Type} & \colhead{A$_{V}$} 
}
\startdata 
IN10-1 &	 18 14 06.343 & -17 28 33.86	& O7.5 V & 7 \\ 
IN10-2 &	 18 14 04.771 & -17 27 58.74	& O6.5 V & 7 \\
IN10-3 &	 18 14 07.104 & -17 29 21.27	& O6 V	 & 5 \\
IN10-4 &	 18 14 06.666 & -17 29 21.34	& O7 V   & 8 \\
\enddata
\tablenotetext{a}{Identification by \citet{watson2008}}
\end{deluxetable}

% Fig 02

Hereafter we will adopt the position of \object{N10} and the ellipse in Figure \ref{map8_objects} as reference: top of the bubble (higher galactic latitude with the center as reference), bottom (lower galactic latitude), right (lower galactic longitude) and left (higher galactic longitude).

%__________________________________________________________________

\section{Observations and data}
\label{data}

%________________________________________________________________

\subsection{CO observations}
\label{CO observations}

The observations were carried out with PMO (Purple Mountain Observatory) 13.7-m radio telescope in 2012 June. We observed the $J=1-0$ transition of $^{12}$CO (115.27 GHz), $^{13}$CO (110.20 GHz) and C$^{18}$O (109.78 GHz). The On-The-Fly (OTF) observing mode was applied to map a $21\arcmin\times25\arcmin$ region centered at $\alpha_{2000}={\rm 18^h14^m01^s.361}$ and $\delta_{2000}=-17\deg 28\arcmin 23\arcsec.14$. For the 13.7-m PMO antenna, we have considered a Half Power Beam Width (HPBW) around $52\arcsec$.

We used a 9 beam array of SIS receivers at the front end \citep{shan2012}. The main beam efficiency at the center of the $3\times3$ array is about 0.44 at 115 GHz and 0.48 at 110 GHz. Our spectral resolution was about 61 kHz, corresponding to velocity resolutions of 0.16 km s$^{-1}$ (at 115 GHz) and  0.17 km s$^{-1}$ (at 110 GHz and 109 GHz).

The cloudy weather condition during our observations led to system temperatures reaching 550 K and 350 K at 115 GHz and 110 GHz, respectively. This resulted in rms noises of 1.7 K and 1.2 K in the brightness temperature for $^{12}$CO $J=1-0$ and $^{13}$CO $J=1-0$, respectively. Such large noise would make relatively weak signals undetectable. However, regions with strong line emission can be validly probed. The velocity information provided by these data convincingly reveal the kinematics of the bubble and molecular conditions in some subregions.

\subsection{Other observations}
\label{Other observations}

\begin{figure*} [ht]
\begin{center}
% \setcaptionmargin{1cm}
\includegraphics[scale=0.29,angle=0]{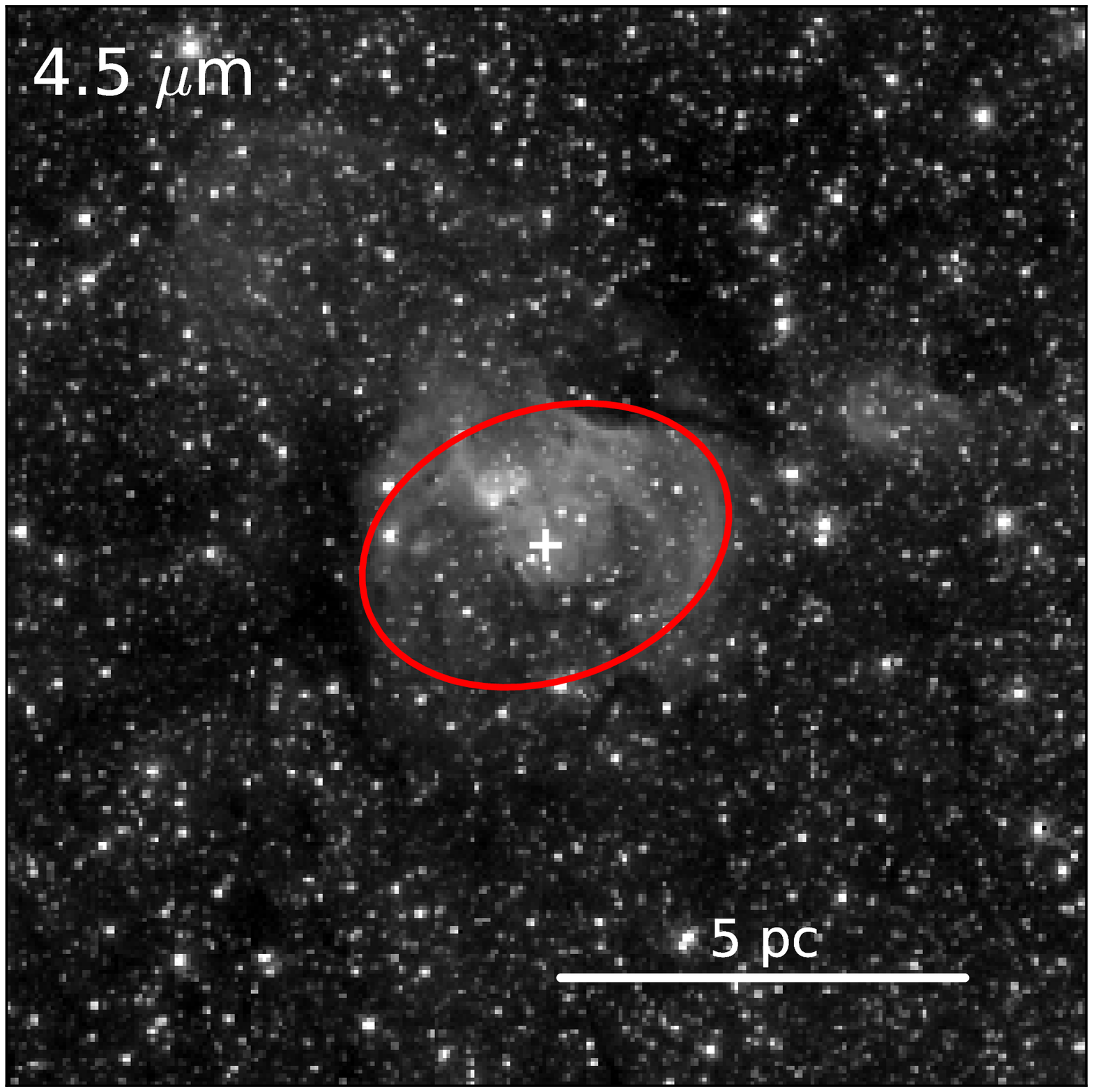}
\includegraphics[scale=0.29,angle=0]{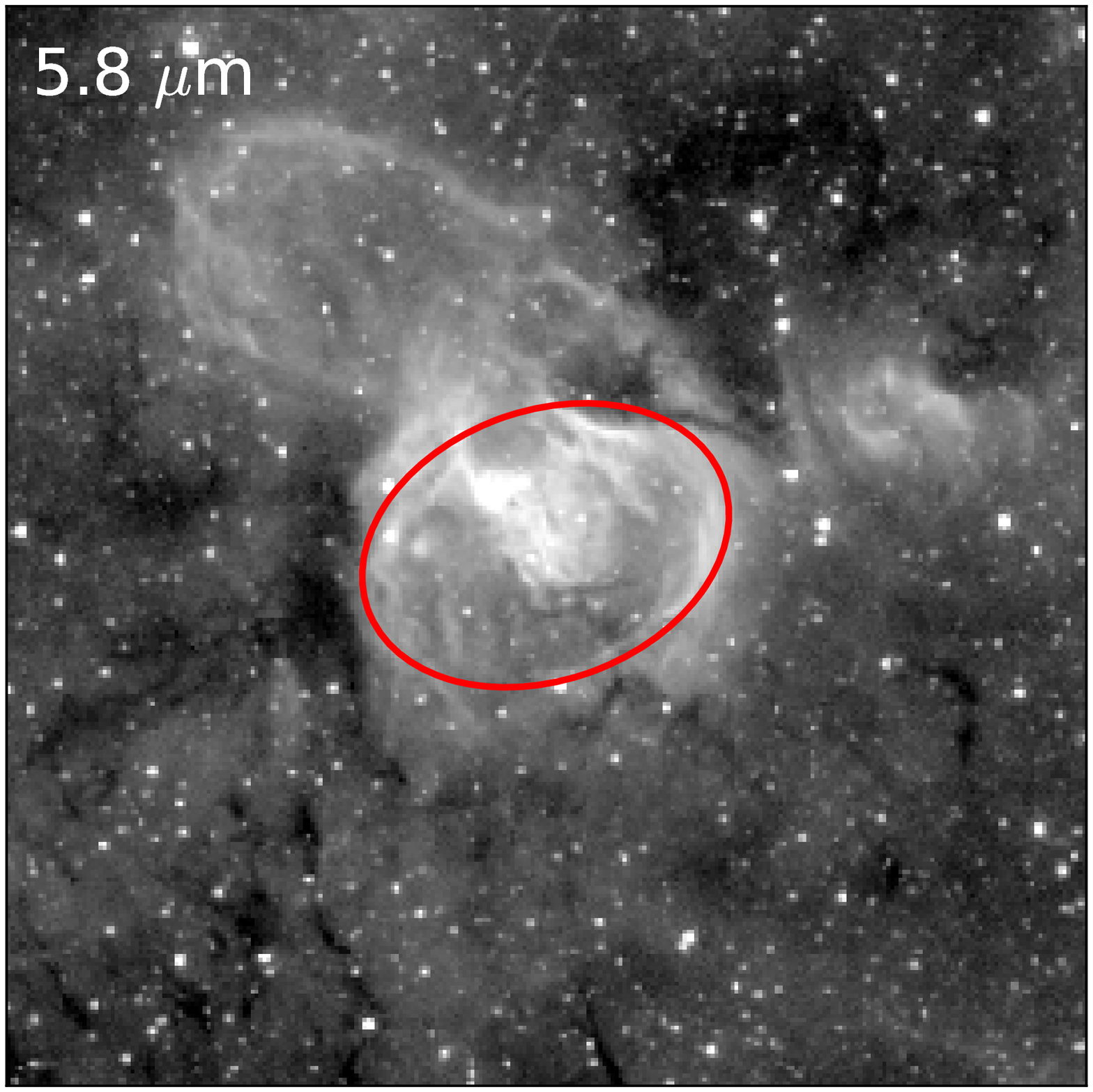}
\includegraphics[scale=0.29,angle=0]{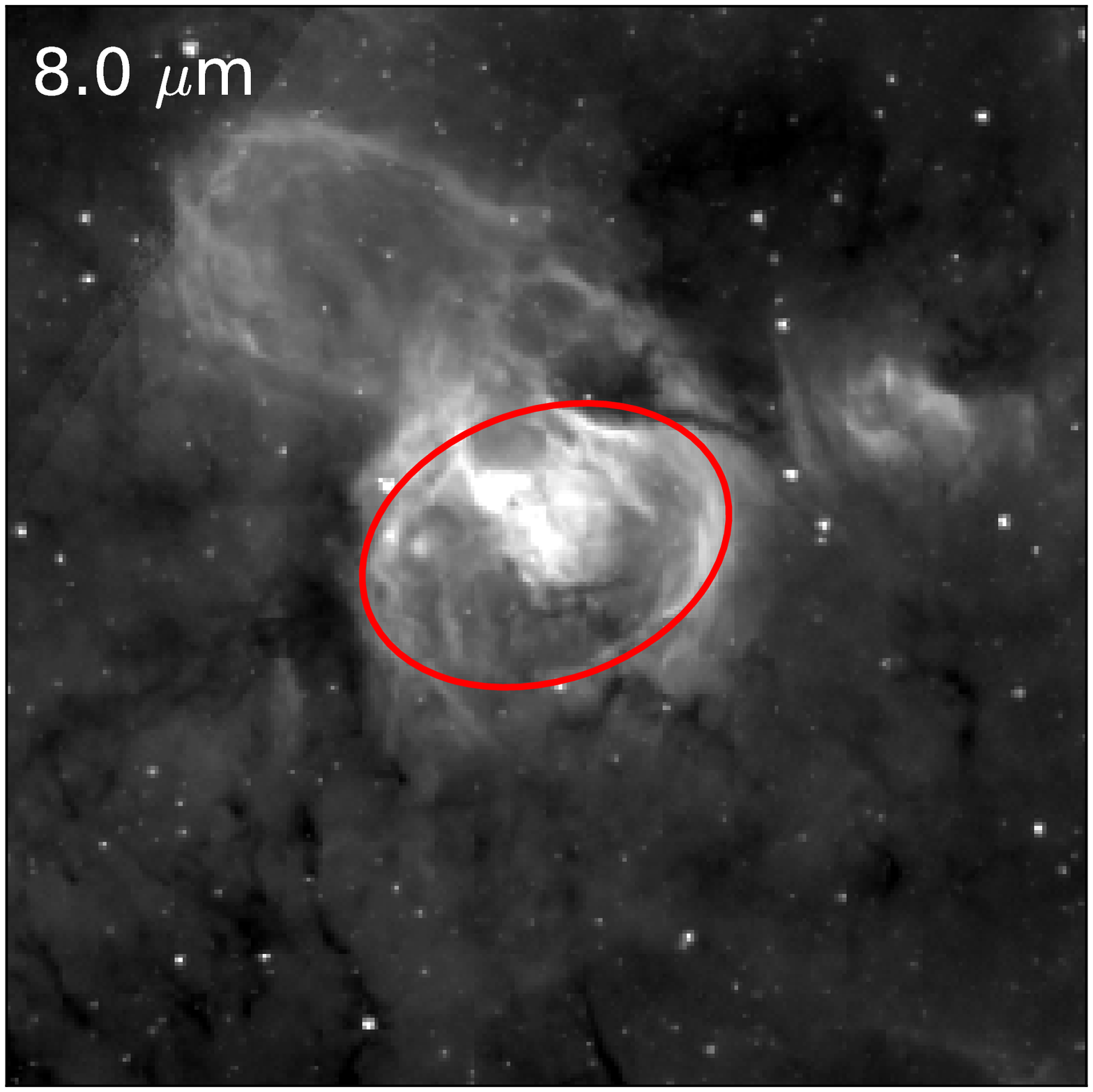}
\includegraphics[scale=0.29,angle=0]{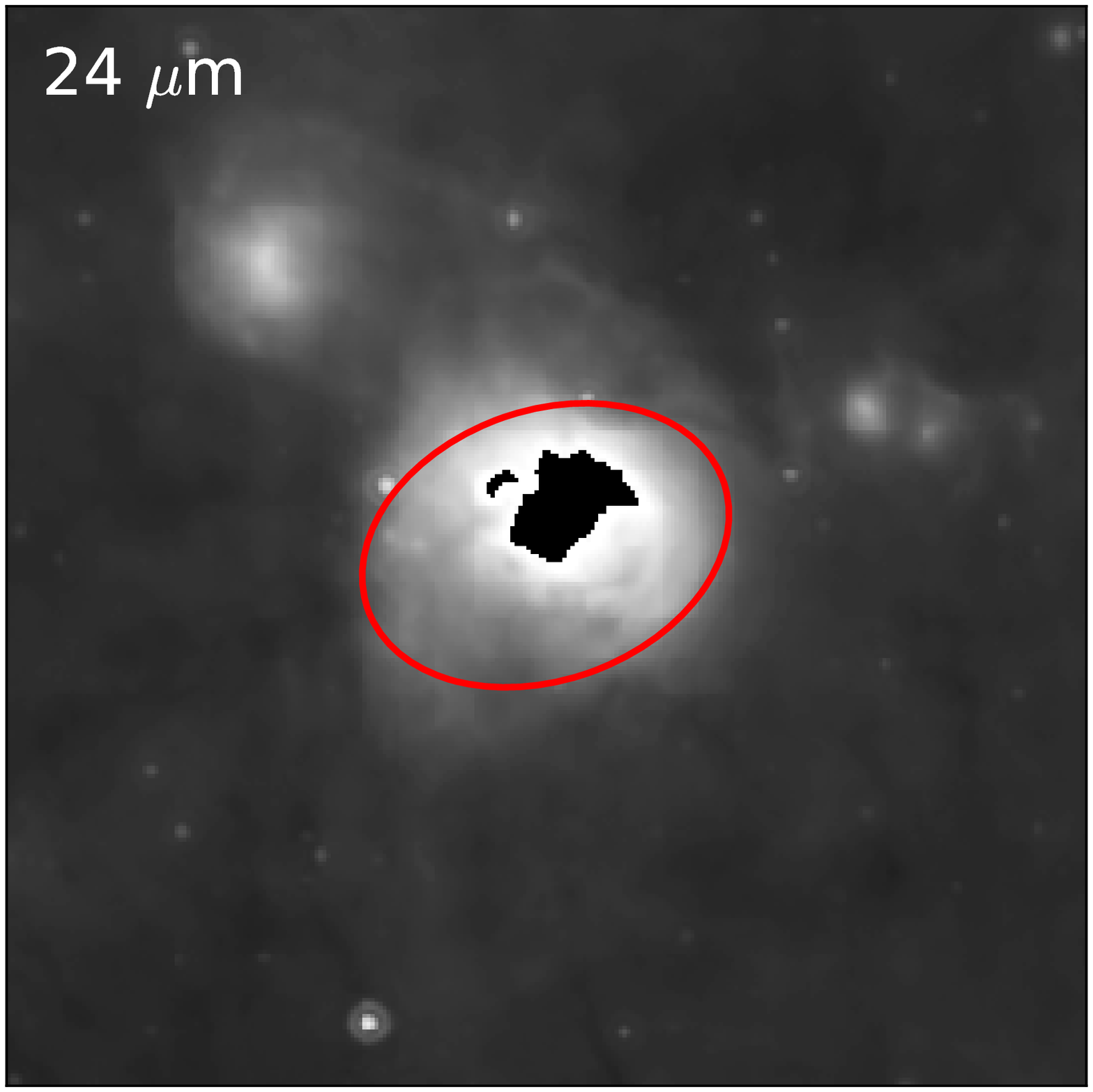}
\includegraphics[scale=0.29,angle=0]{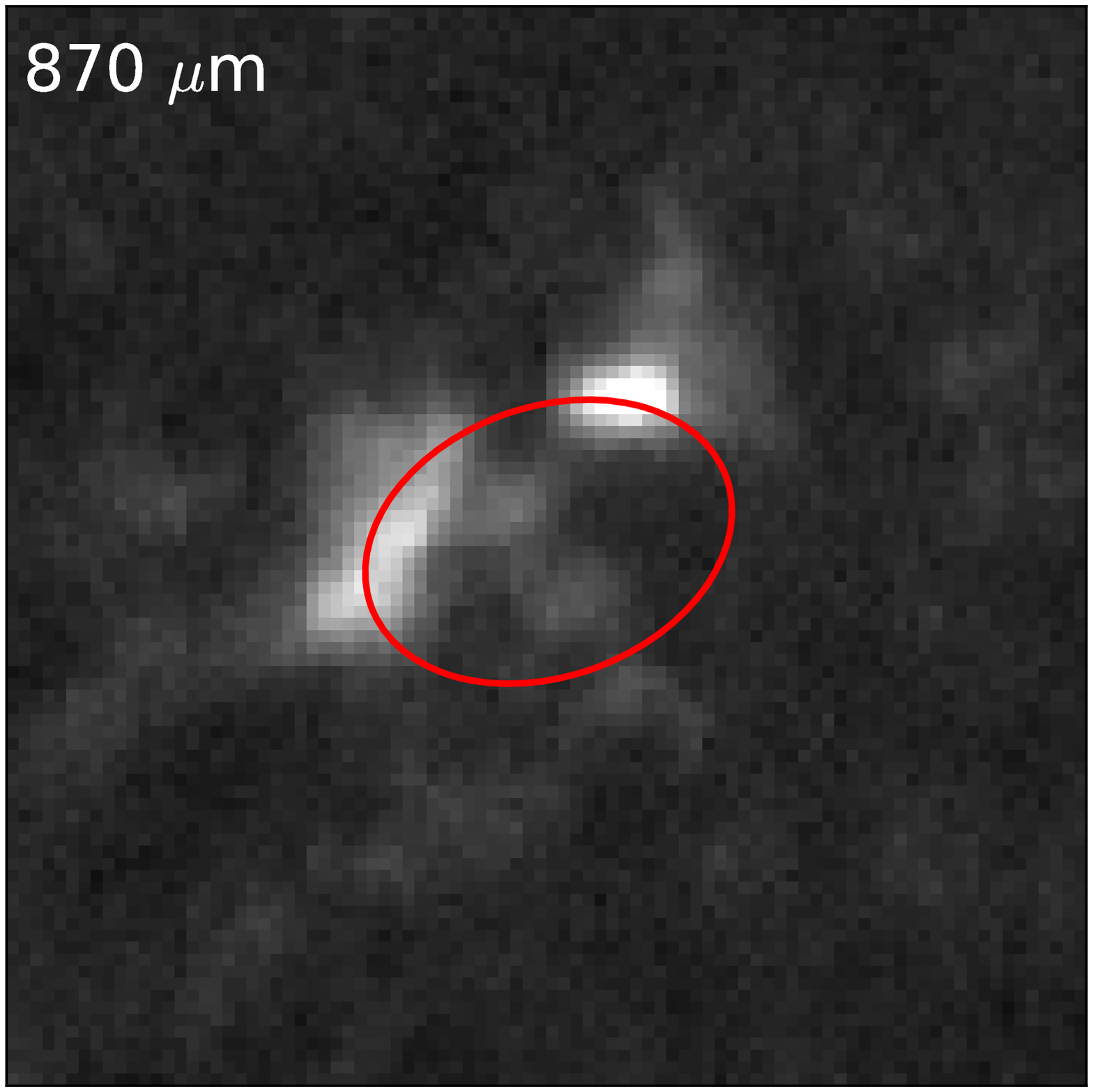}
\includegraphics[scale=0.29,angle=0]{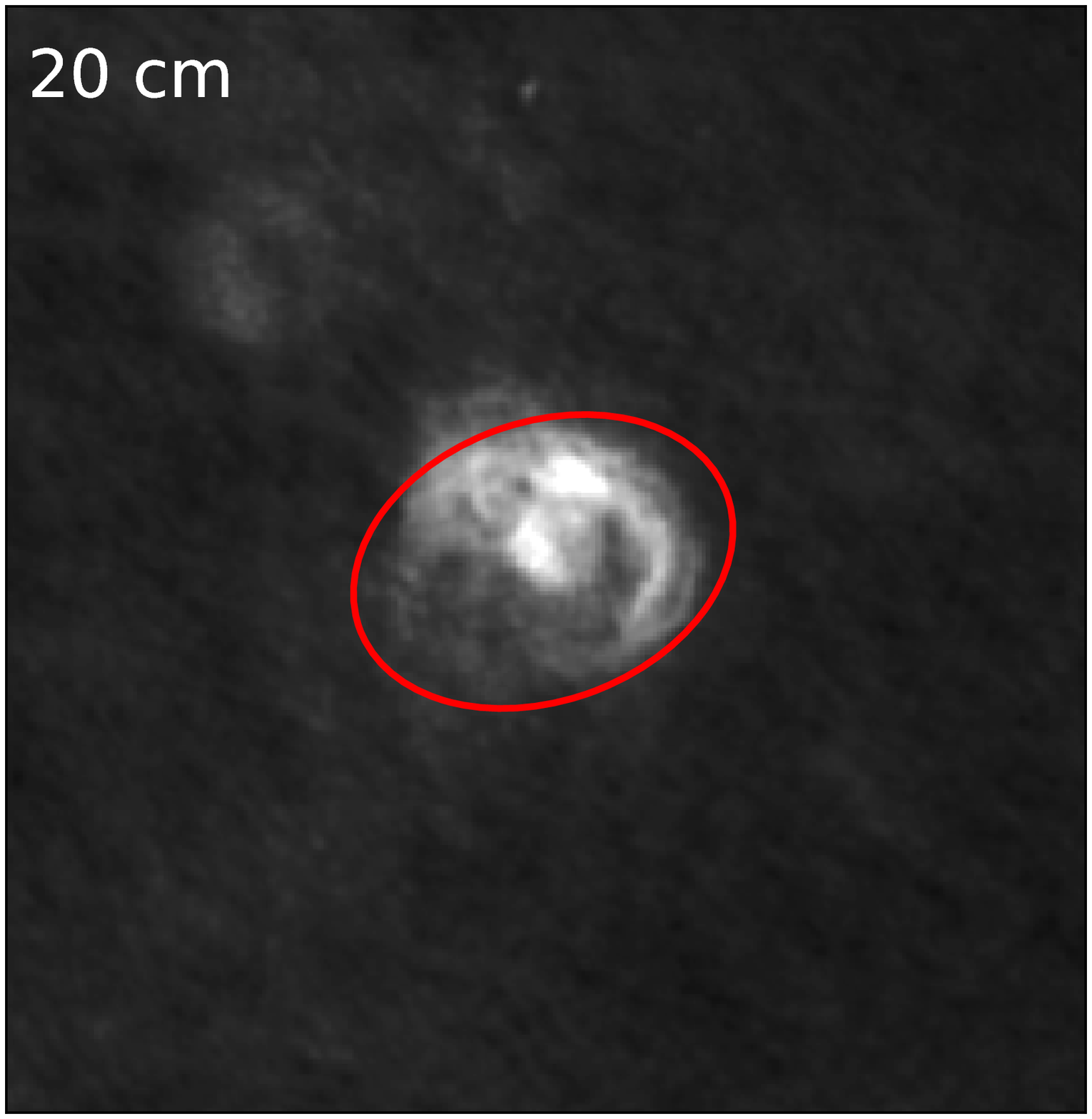}
\caption{Multi-wavelength images of the N10 region. The panels show 4.5 $\mu$m, 5.8 $\mu$m, 8.0 $\mu$m, 24 $\mu$m, 870 $\mu$m and 20 cm emission towards the target region, from Spitzer-GLIMPSE, Spitzer-MIPSGAL, LABOCA-ATLASGAL and GPS-VLA, respectively. All the images are showing the same region in the sky; in the first panel, white cross marks the center of the HII region; in all images the red ellipse is for reference, as shown in Figure \ref{n10_8microns}.} 
\label{multi_wave}
\end{center}
\end{figure*}

Public data from infrared to centimeter surveys was used to analyze the bubble \object{N10} at other wavelengths. The GLIMPSE survey \citep{benjamin2003} mapped parts of the inner Galactic plane, with IRAC \citep[Infrared Array Camera;][]{fazio2004} on Spitzer Space Telescope. We obtained images of 4.5, 5.8 and 8.0 $\mu$m IRAC bands Figure \ref{multi_wave}. The 24 $\mu$m image of N10 was obtained from another survey of the inner Galactic plane, MIPSGAL, using the MIPS instrument \citep[Multiband Imaging Photometer for the Spitzer; ][]{rieke2004}. In the panel at 24 $\mu$m of Figure \ref{multi_wave} we can see that this emission fills the whole area indicated by the red ellipse. This emission, typical of galactic bubbles, is caused by warm dust present in the region of ionized gas. 

% Fig 03

The existence of an HII region inside the bubble is confirmed by the radio continuum emission at 20 cm from MAGPIS \citep{helfand2006}. In order to improve the continuum emission data, MAGPIS combined VLA images with images from a 1.4 GHz survey carried out by \cite{reich1990} using the Effelsberg 100-m telescope. This emission, due to free-free process, is a good tracer of ionized gas.

\object{N10} was mapped at sub-mm wavelengths with the APEX telescope \citep{miettinen2012}. The images at 870 $\mu$m wavelength were obtained with ATLASGAL (APEX Telescope Large Area Survey of the Galaxy), an observing program using the LABOCA \citep[Large Apex BOlometer CAmera instrument][]{schuller2009}. The 870 $\mu$m cold dust emission is useful to reveal the presence of dense dark clouds. Two of these clouds, reported by \citet{wienen2012}, are seen bordering the bubble, along the upper and left borders of the bubble. The HII region, which is probably expanding, seems to be interacting with these clouds. We used the MAGPIS website\footnote{http://third.ucllnl.org/gps} to obtain the images presented Figure \ref{multi_wave}.

We also used the all-sky Wide-Field Infrared Survey Explorer satellite \citep[WISE;][]{wright2010} data to analyze the content of young stellar objects in \object{N10}, in order to reveal the regions where star formation took place recently and possible gradients of evolutionary stage. 

%__________________________________________________________________

\section{Results}
\label{results}

%________________________________________________________________

\subsection{Molecular Emission}
\label{Molecular Emission}

\begin{figure*}[ht]
\begin{center}
% \setcaptionmargin{1cm}
%\includegraphics[scale=0.4,angle=0]{fig04.eps}
\includegraphics[scale=0.55,angle=0]{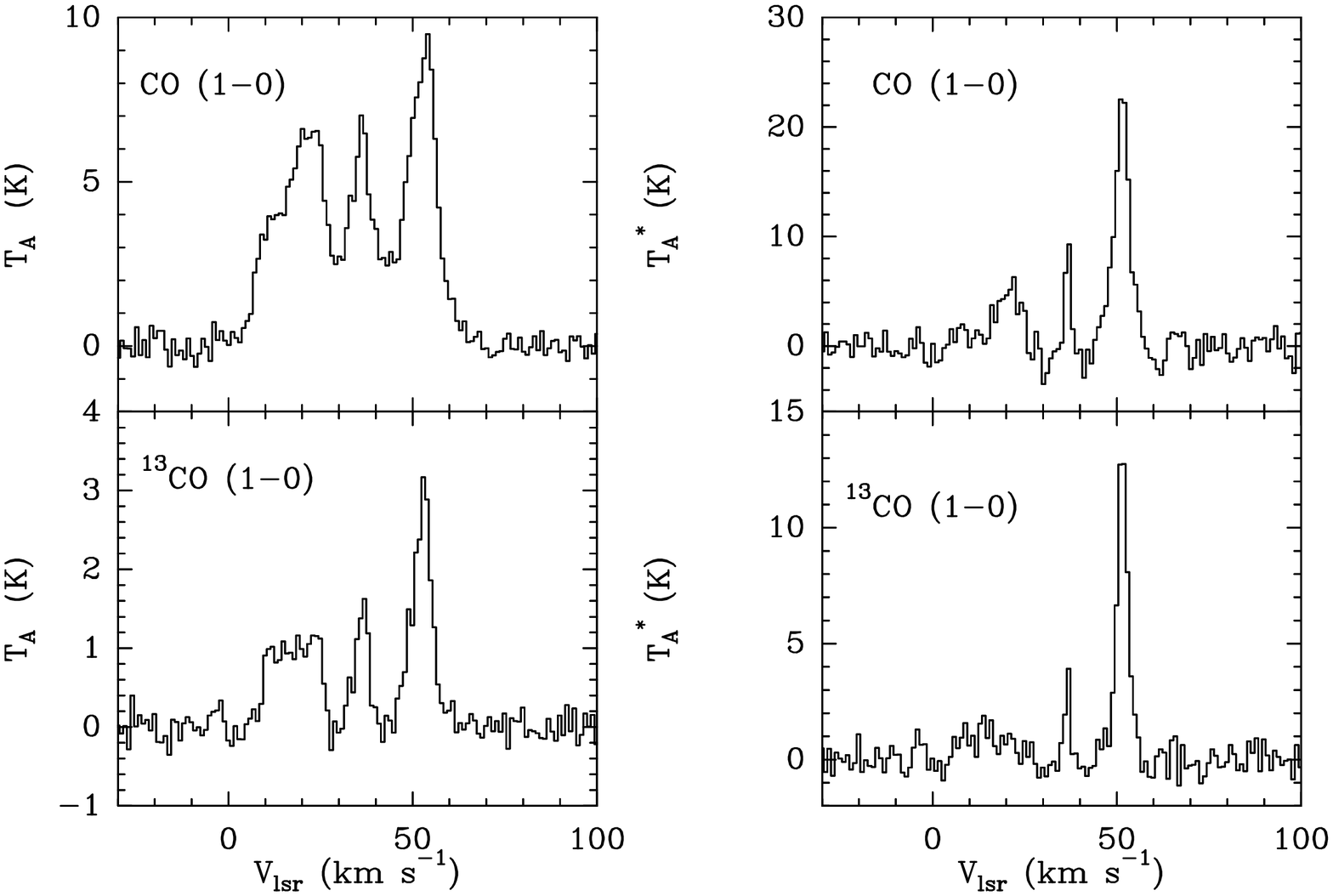}
\caption {Spectral lines. The left column shows the average spectra of $J=1-0$ transition of $^{12}$CO in the upper panel and $^{13}$CO in the lower panel of the observed region ($\Delta l = 0.002^{\circ}$, $\Delta b = 0.304^{\circ}$). The right column presents spectra of these two lines in the peak position, at $l = 13.21^{\circ}$, $b = 0.037^{\circ}$.}
\label{spectra}
\end{center}
\end{figure*}

The emission of $^{12}$CO, $^{13}$CO and C$^{18}$O J=1-0 was observed at the same time. Strong emission of $^{12}$CO and $^{13}$CO was observed; the emission of C$^{18}$O is weak and we do not analyze in this work. The Figure \ref{spectra} presents the observed spectral lines. 

% Fig 04

\begin{figure*}
\begin{center}
% \setcaptionmargin{1cm}
\includegraphics[scale=0.42,angle=0]{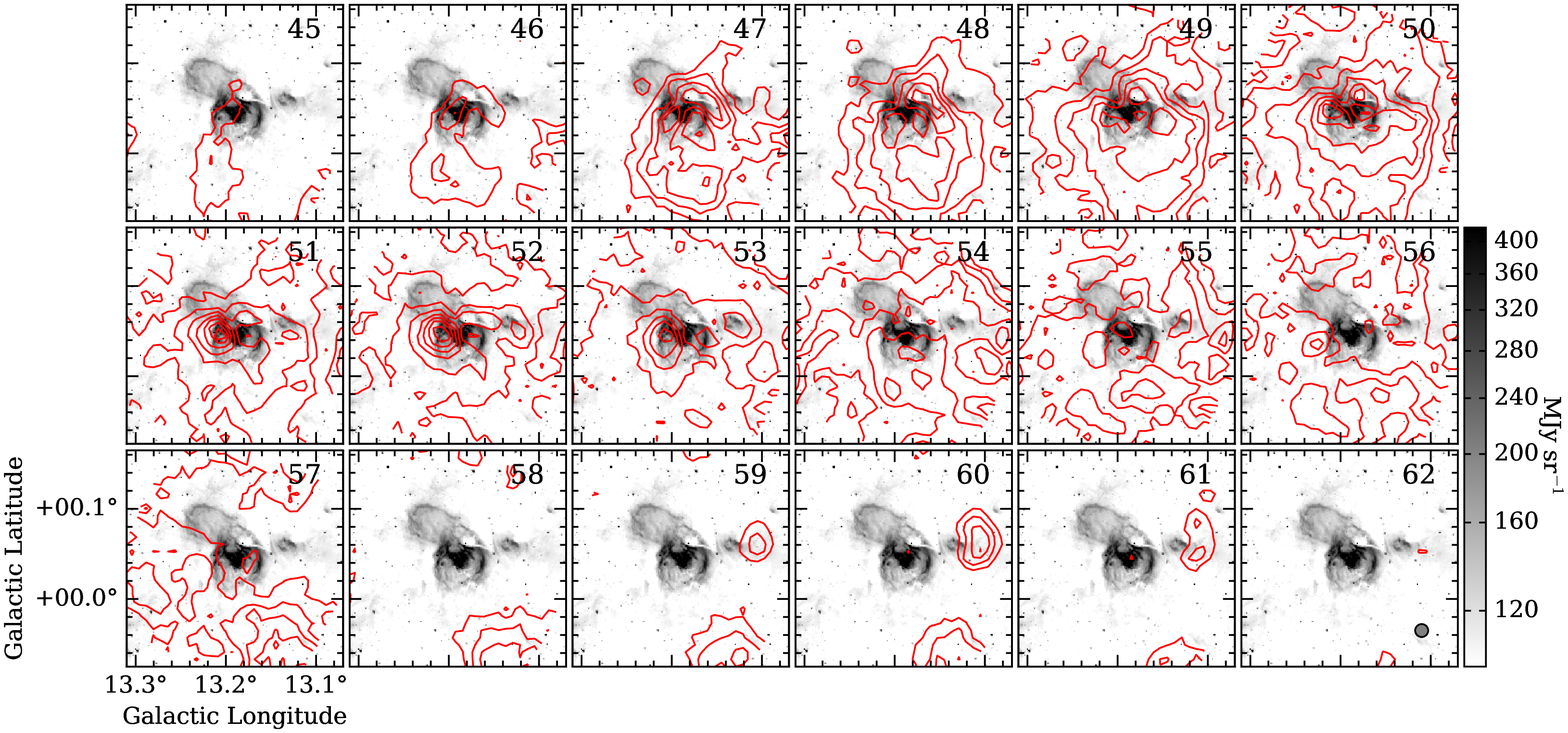}\\
\includegraphics[scale=0.42,angle=0]{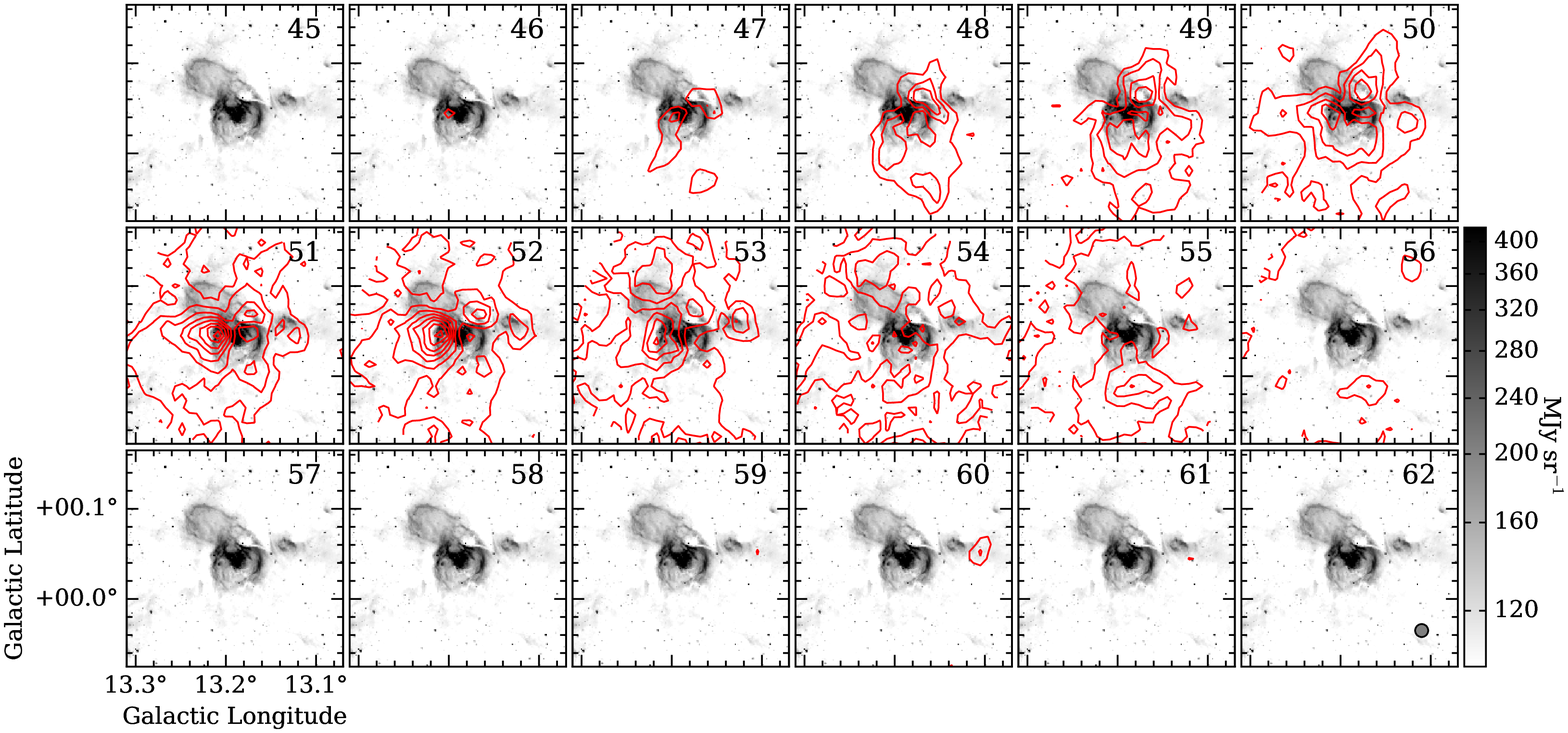}
\caption {\textit{Upper panel}: Channel maps of $^{12}$CO J = 1-0 emission in contours superimposed on the Spitzer 8 $\mu$m image. The contours start from 5 $\sigma$ with steps of 3 $\sigma$ (1 $\sigma$ = 0.90 K km $s^{-1}$). The main velocity in km s$^{-1}$ is indicated in the left-upper corner of each panel.
%, velocity interval of integration covering 1 km s$^{-1}$ with the center value. 
The small circle represents the beam size of $^{12}$CO observations. \textit{Lower panel}: The same, for $^{13}$CO J = 1-0 emission. The contours start from 5 $\sigma$ increasing with a step of 3 $\sigma$ (1 $\sigma$ = 0.55 K km s$^{-1}$). The scale bar shows the 8 $\mu$ flux intensity on a logarithmic scale and the small circle represents the beam size of $^{13}$CO observations.}
\label{channel_map}
\end{center}
\end{figure*}

Detected $^{12}$CO and $^{13}$CO emission allows us to identify three peaks of velocity: at 20, 37 and 52 km s$^{-1}$, approximately. The central velocities and line widths were determined by Gaussian fits using the CLASS package (GILDAS software\footnote{http://www.iram.fr/IRAMFR/GILDAS}). In this paper, velocities are referred to the local standard of rest (V$_{LSR}$). Upper panel in Figure \ref{channel_map} displays a channel map of $^{12}$CO emission and bottom panel shows the channel map of $^{13}$CO emission. The background in both figures shows 8.0 $\mu$m emission. We have fitted the channels by increasing the velocity from 45 to 62 km s$^{-1}$. There is an strong correlation between $^{12}$CO and $^{13}$CO emission, specially in the range 48-53 km s$^{-1}$. 

% Fig 05

Broad CO component centered at 20 km s$^{-1}$ has the lower intensity of the three peaks. Component centered at 37 km s$^{-1}$ presents narrower profile than the former, and lower intensity if compared with the component centered at 52 km s$^{-1}$. 

The velocities found in the literature for different emission lines associated with N10 range from 48.5 to 54.1 km s$^{-1}$, as shown in Table 2. This leads us to adopt the component with peak at 52.6 km s$^{-1}$ as the one related to the source.

% Table 02
\begin{deluxetable}{c c c} 
%\tabletypesize{\scriptsize}
%\tablecolumns{3} 
\tablewidth{0pt} 
\tablecaption{Velocities derived towards the bubble N10.\label{velocities}} 
\tablehead{ 
\colhead{Velocity} & \colhead{Method} & \colhead{Reference}
}
\startdata 
54.1 km s$^{-1}$ & H II region, radio recombination line & 1 \\
48.5 km s$^{-1}$ & 6.7 GHz methanol maser emission & 2  \\
54.1 km s$^{-1}$ & HI absorption line & 3 \\
50.2 km s$^{-1}$ & CO line emission & 4 \\
48.5 km s$^{-1}$ & NH$_{3}$ inversion line (from 870 $\mu$m data) & 5 \\
54.1 km s$^{-1}$ & mid-infrared from WISE HII region & 6 \\
\enddata
%\tablenotetext{a}{Identification by \citet{watson2008}}
\tablerefs{(1) \citealp{lockman1989}; (2) \citealp{szymczak2000}; (3) \citealp{pandian2008}; (4) \citealp{beaumont2010}; (5) \citealp{wienen2012}; (6) \citealp{anderson2014}.}
\end{deluxetable}

\begin{figure*}
\begin{center}
% \setcaptionmargin{1cm}
\includegraphics[scale=0.28,angle=0]{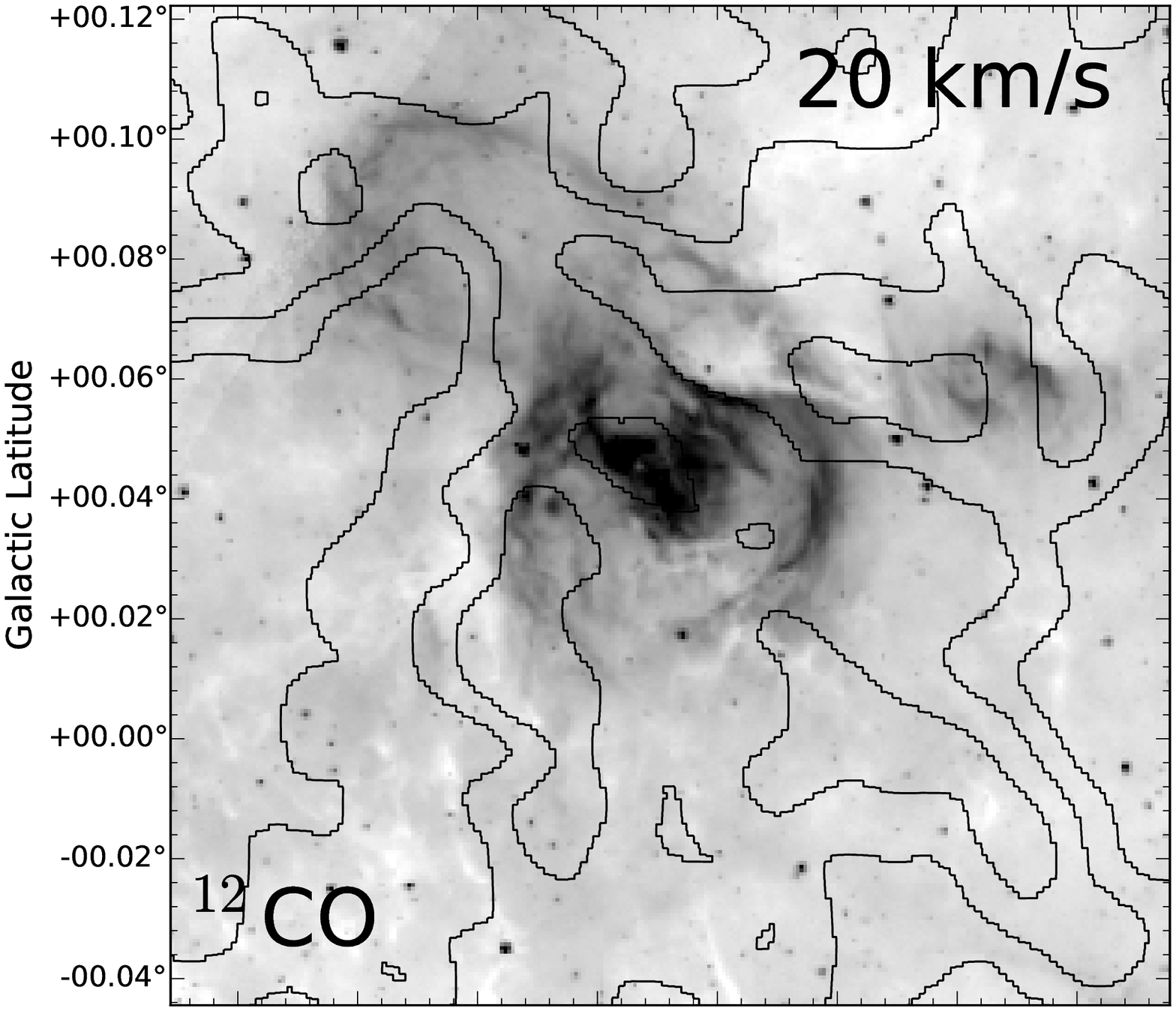}
\includegraphics[scale=0.28,angle=0]{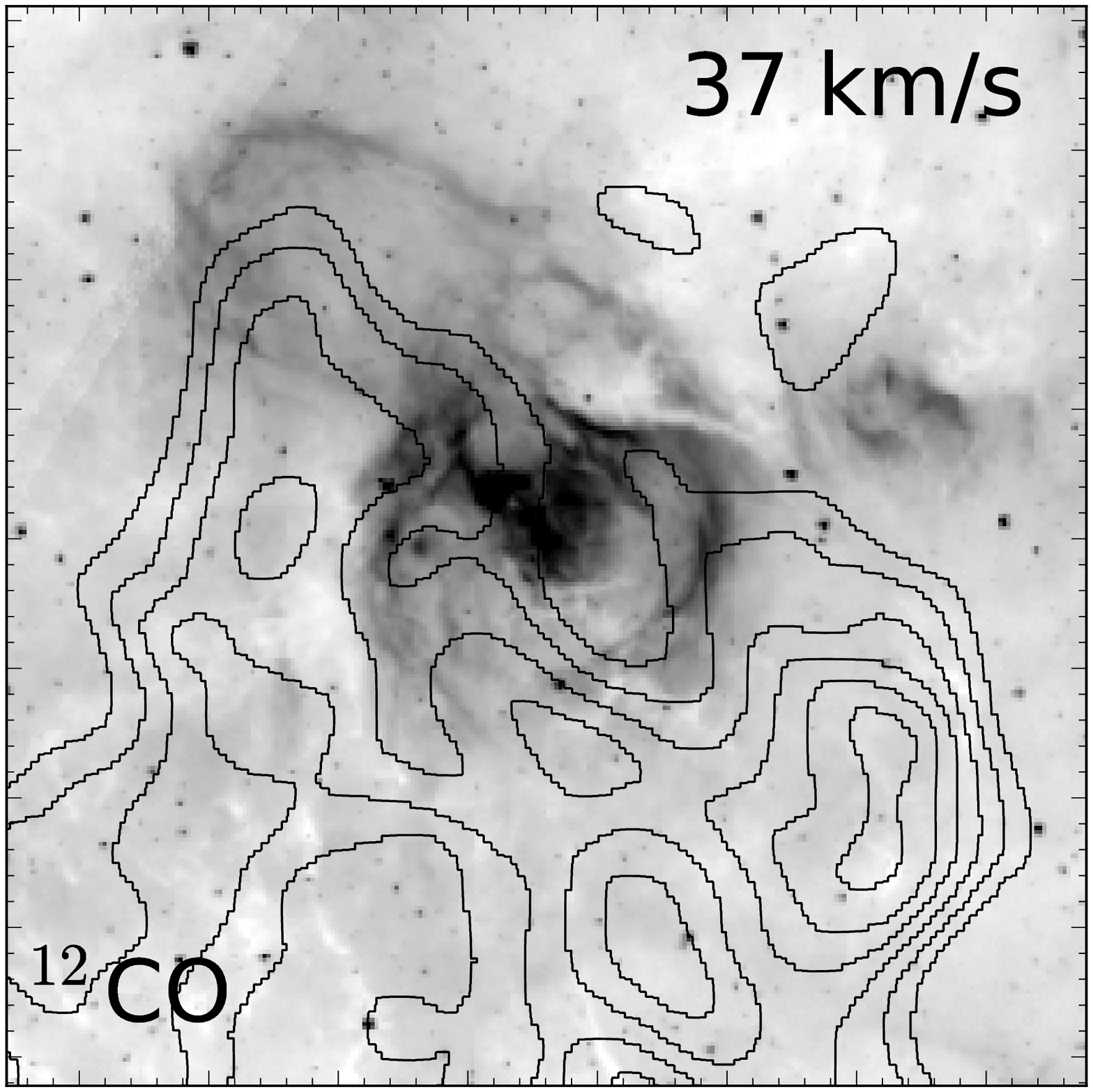}
\includegraphics[scale=0.28,angle=0]{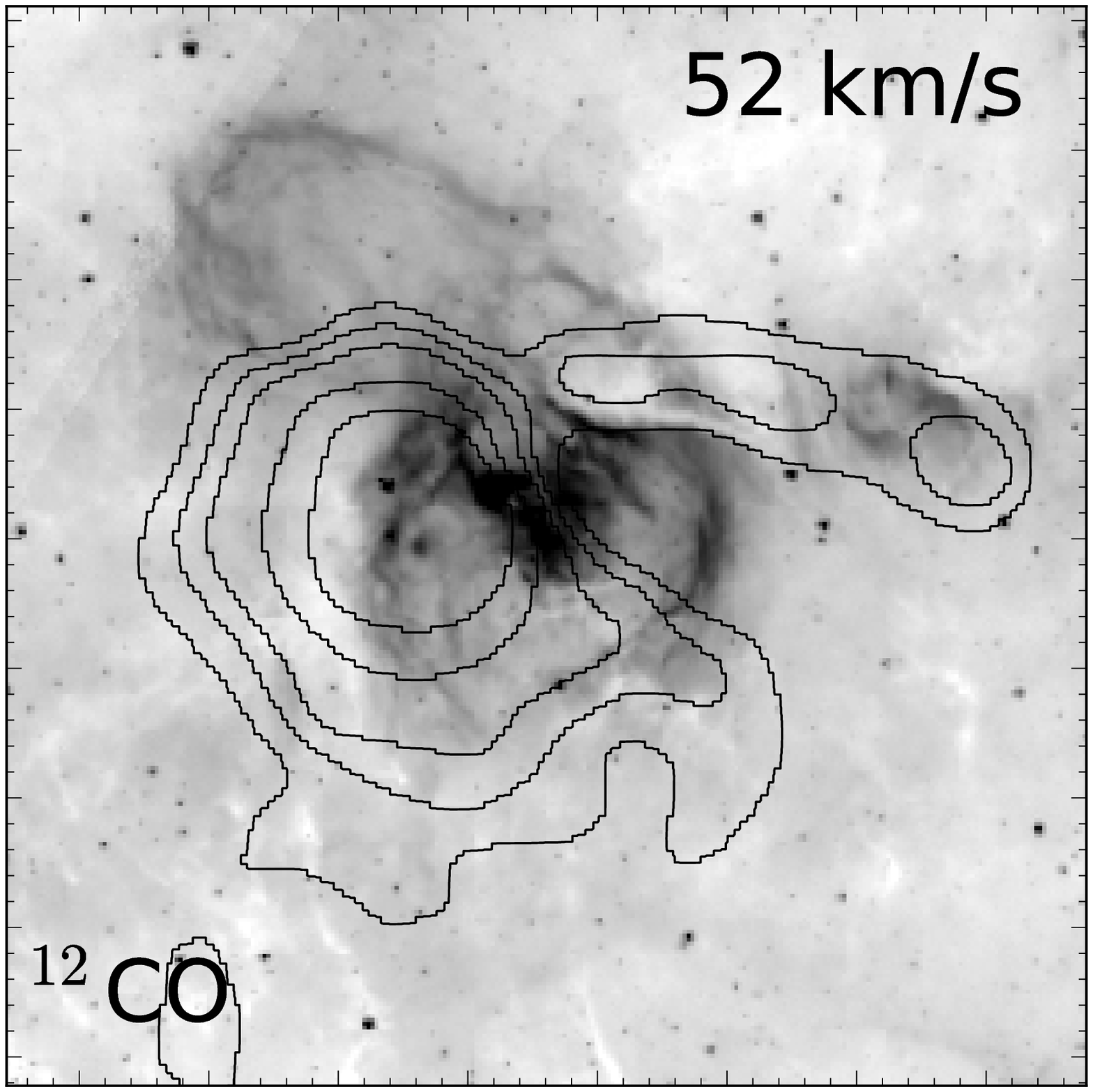}
\includegraphics[scale=0.28,angle=0]{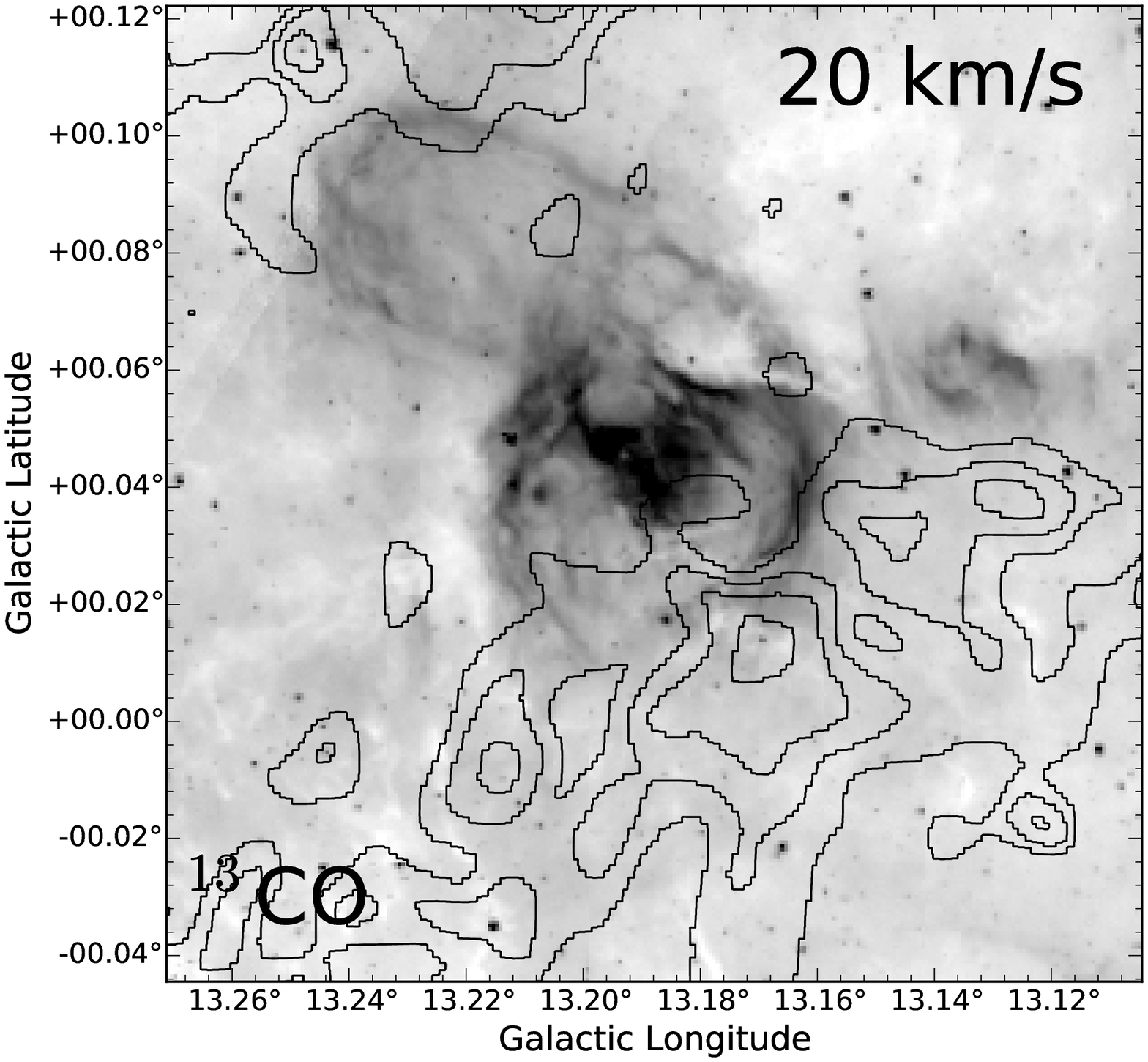}
\includegraphics[scale=0.28,angle=0]{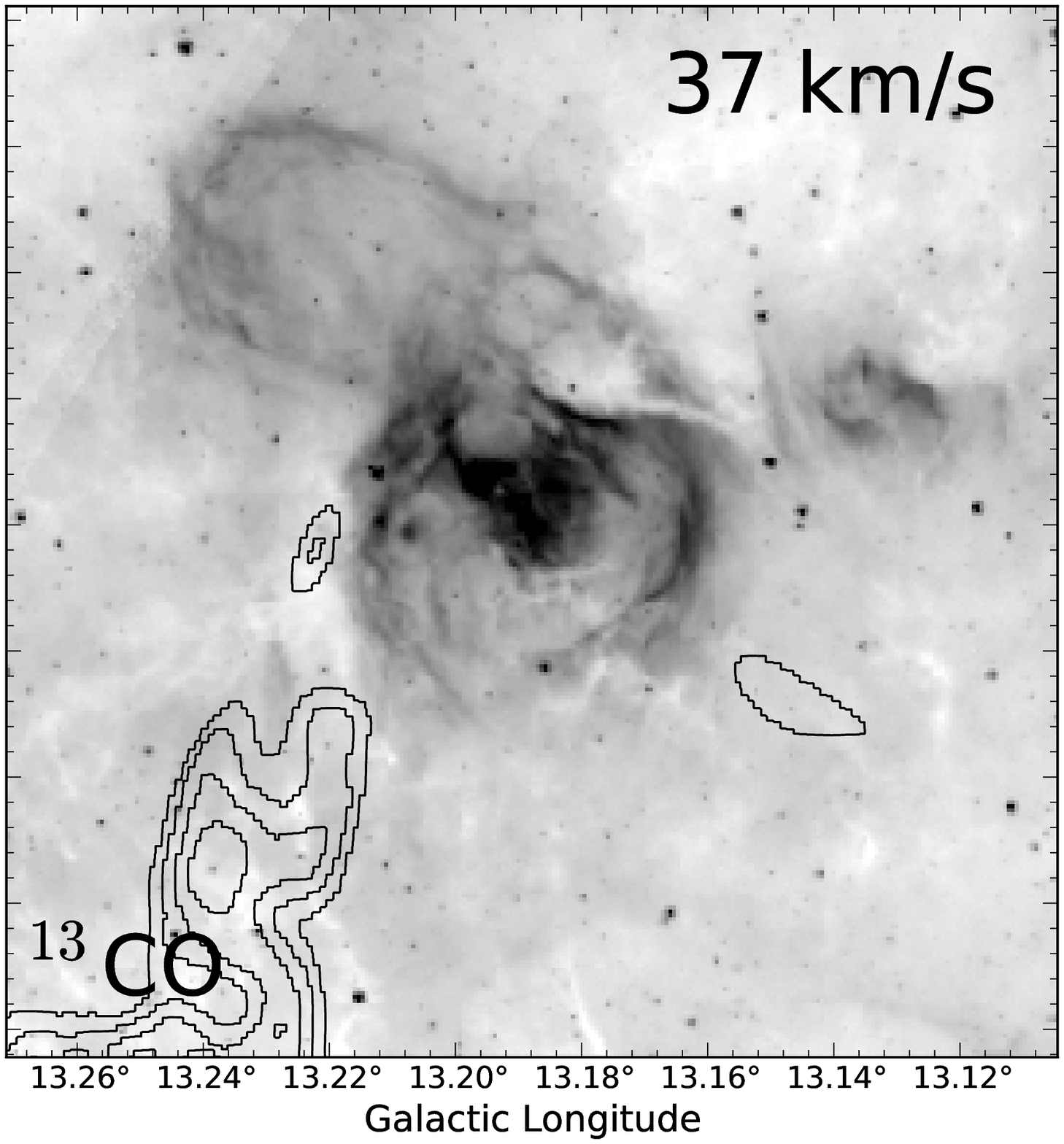}
\includegraphics[scale=0.28,angle=0]{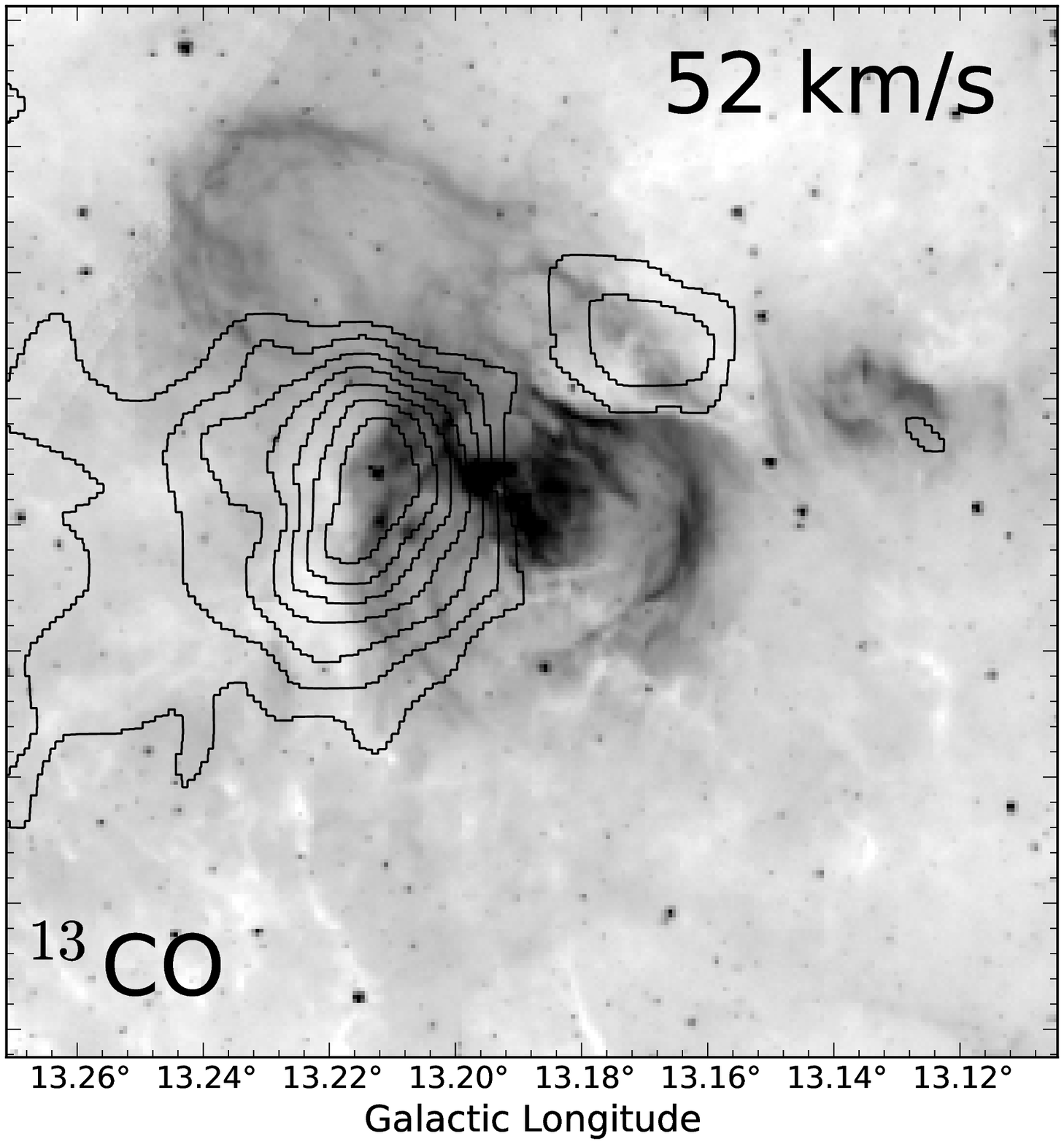}
\caption{The upper panels show contours of $^{12}$CO emission distribution centered at 20 km s$^{-1}$ from 6 to 13-K in steps of 1-K (\textit{left}); at 37 km s$^{-1}$ in steps of 1-K, from 8 to 14-K (\textit{middle}); at 52 km s$^{-1}$ from 5 to 12-K, in steps of 1-K (\textit{right}). 
The lower panels display contours of $^{13}$CO emission distribution centered at at 20 km s$^{-1}$ from 1 to 3-K in steps of 0.5-K (\textit{left}); at 37 km s$^{-1}$ in steps of 1-K, from 5 to 10-K (\textit{middle}); at 52 km s$^{-1}$ from 10 to 21-K, in steps of 1-K (\textit{right}). All panels exhibit 8.0 $\mu$m image in background.} 
\label{8micron_co}
\end{center}
\end{figure*}

In our observation, velocities along the emission with peak at 52.6 km s$^{-1}$ range from 48 to 53 km s$^{-1}$. In order to verify the correspondence between the physical distribution of molecular gas and the bubble seen in infrared, $^{12}$CO and $^{13}$CO contours of narrow-velocity emission were superposed over a Spitzer 8.0 $\mu$m image in Figure \ref{8micron_co}. Spatial distribution of $^{12}$CO shows two main structures that seem to be related to the 8.0 $\mu$m emission, and $^{13}$CO presents two denser clumps in the border of the ring morphology of \object{N10}, at the same position of $^{12}$CO structures.

% Fig 06

\subsection{Distribution of gas and dust}
\label{Distribution of gas and dust}

We have studied the bubble \object{N10} through the emission of the CO and the cold dust, which is useful to reveal the densest and coldest regions of \object{N10}. However, it is necessary to explore other tracers.

\subsubsection*{Ionized gas}

Ionized gas associated with \object{N10} can be traced by VLA 20 cm emission. The presence of emission at $\nu = 1.5$ GHz implies that the HII region in the inner part of the bubble is created by UV photons. Using the Greg/GILDAS software we estimated the 20 cm total flux $F_{20\ cm}=1.17$ Jy inside the bubble and we calculated the electron density ($n_{e}$) according \cite{panagia1978}: 

\begin{eqnarray} \label{panagia}
\frac{\mathrm n_{e}}{\mathrm cm^{-3}} &=& \ 3.113 \times 10^{2}  
\Bigg( \frac{\mathrm  T_{e}}{\mathrm  10^{4} \ K} \Bigg)^{0.25} 
\Bigg( \frac{\mathrm  S_{\nu}}{\mathrm Jy} \Bigg)^{0.5} 
\Bigg( \frac{\mathrm D}{\mathrm kpc} \Bigg)^{-0.5} \nonumber
\\
& \times & b({\mathrm \nu}, {\mathrm T})^{-0.5} \times 
\theta_{\mathrm R}^{-1.5},
\end{eqnarray}

\noindent where ${\mathrm T_{e}}$ in K is the electron temperature, ${\mathrm S_{\nu}}$ is the measured total flux density in Jy, ${\mathrm D}$ is the distance in kpc and ${\mathrm \theta_{R}}$ is the angular radius in arcmin. The function ${\mathrm b(\nu, T)}$ is defined as:

\begin{eqnarray} \label{b-funtion}
b({\mathrm \nu}, {\mathrm T}) \ &=& \ 1 \ 
+ \ 0.3195 \ 
log \ \Bigg( \frac{\mathrm T_{e}}{\mathrm 10^{4} \ K} \Bigg) \nonumber
\\
&-& \ 0.2130 \ 
log \ \Bigg( \frac{\mathrm \nu}{\mathrm 1 \ GHz} \Bigg).
\end{eqnarray}

Assuming ${\mathrm T_{e}=10^{4} \ K}$ for the free-free emission region, the electron density is ${\mathrm n_{e} = 129.71}$ cm$^{-3}$.

\begin{figure*}
\begin{center}
\includegraphics[scale=0.42,angle=0]{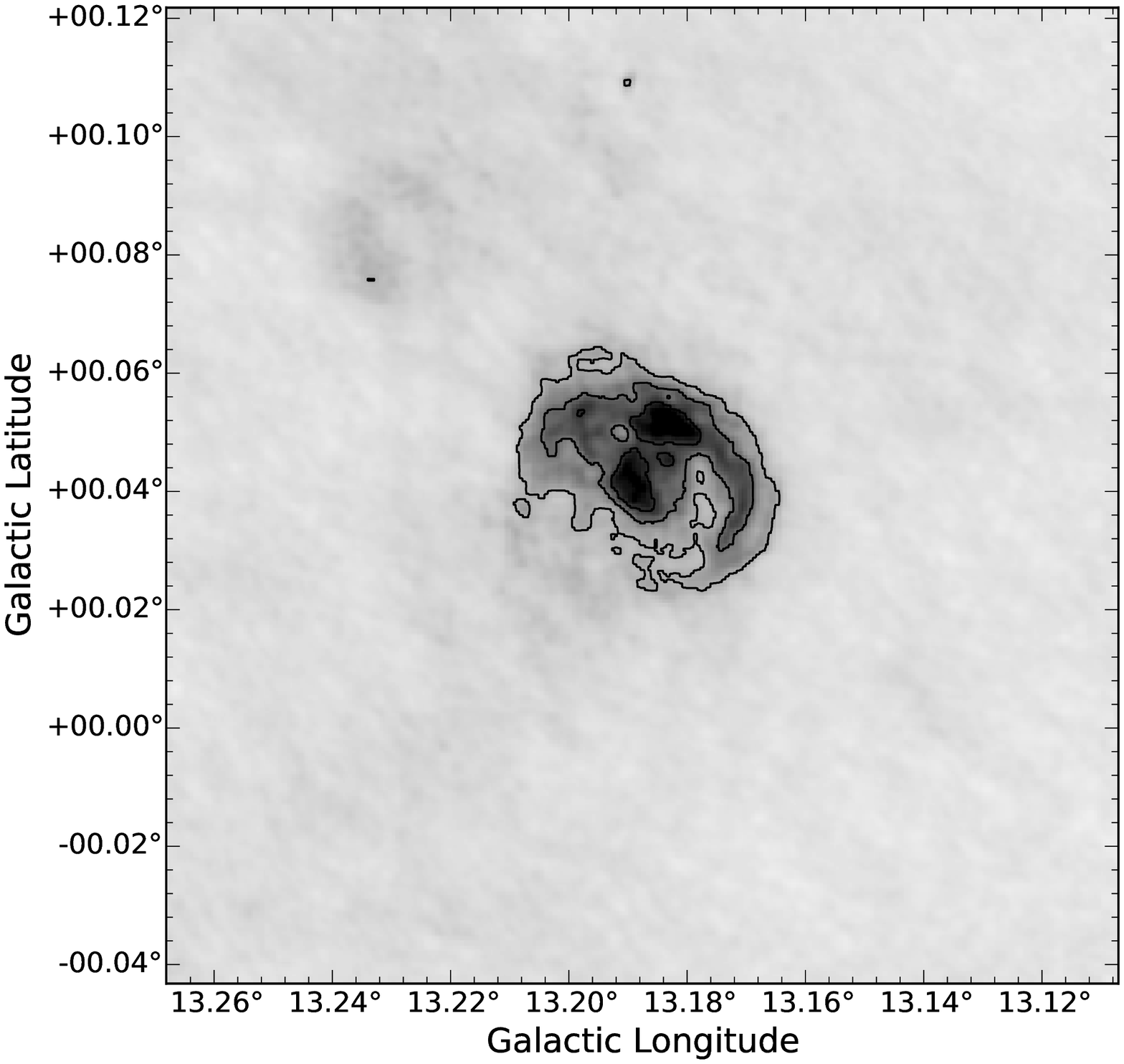}
\includegraphics[scale=0.42,angle=0]{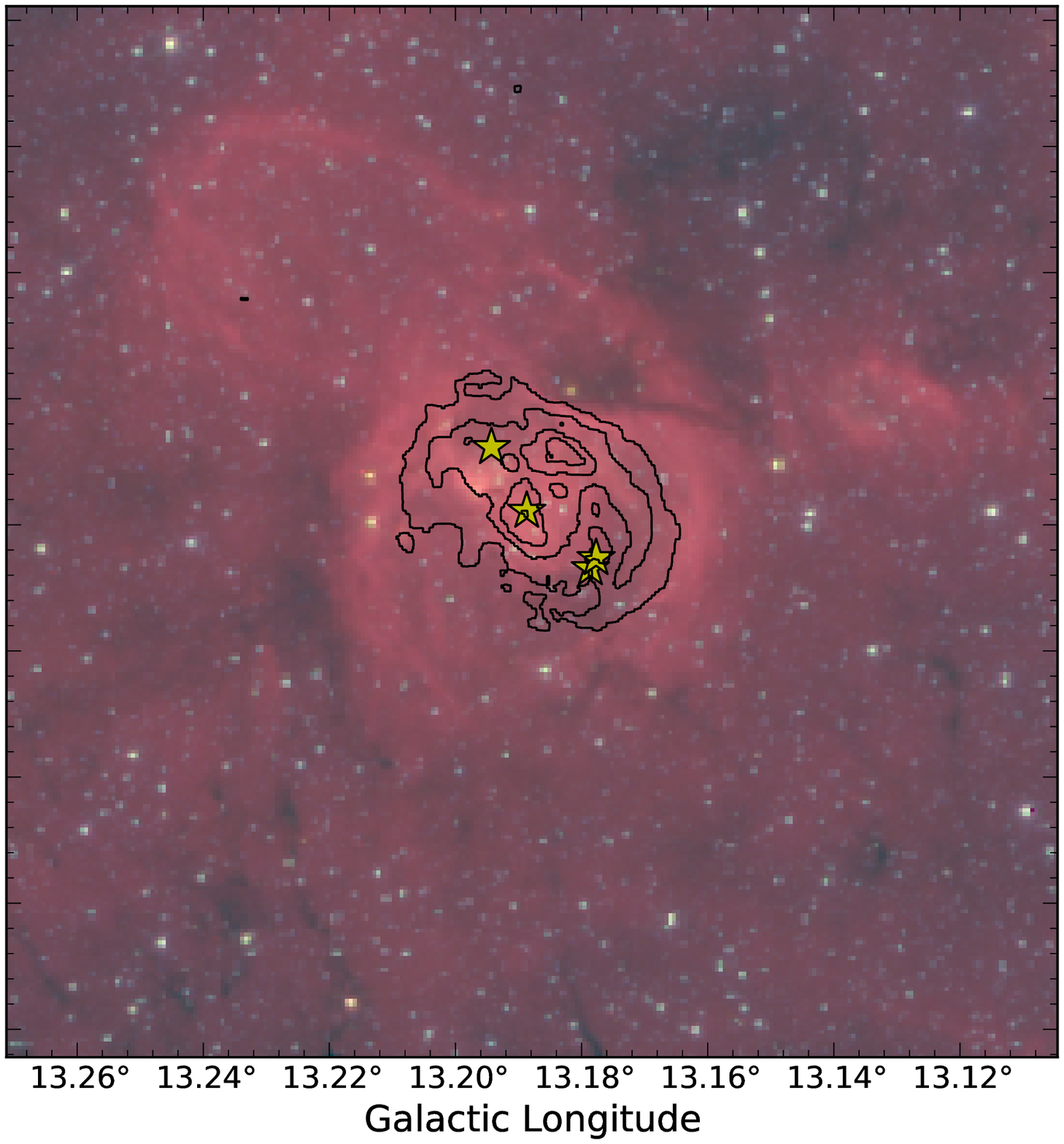}
\caption{\textit{Left panel}: VLA 20 cm (1.4 GHz) intensity in grayscale, from 0 to 20 mJy beam$^{-1}$, with contours start from 2 $\sigma$ with steps of 2 $\sigma$ (1 $\sigma$ = 2.8 mJy beam$^{-1}$). \textit{Right panel}: RGB Spitzer image in background (3.6 $\mu$m in blue, 4.5 $\mu$m in green and 8.0 $\mu$m in red). Yellow stars marks the position of the candidates ionizing stars.}
\label{map8_cont20}
\end{center}
\end{figure*}

Figure \ref{map8_cont20} displays two peaks of radio continuum emission in grayscale and black contours in left panel. In the same figure, right panel shows one of the peaks coinciding with an O-type star.

The number of Lyman continuum photons that are absorbed by the gas in the region HII was calculated using the radio continuum map, following the relation given by \citet{matsakis1976}:

\begin{eqnarray} \label{lyman}
\mathrm N_{uv} = 
7.5 \times 10^{46}
\Bigg( \frac{\mathrm \nu}{\mathrm GHz} \Bigg)^{0.1}
\Bigg( \frac{\mathrm T_{e}}{\mathrm 10^{4} \ K} \Bigg)^{-0.45}
\Bigg( \frac{\mathrm S_{\nu}}{\mathrm Jy} \Bigg) \ 
\Bigg( \frac{\mathrm D}{\mathrm kpc} \Bigg)^{2}
\end{eqnarray}

We estimate ${\mathrm N_{uv}=1.86 \times 10^{49}}$ ionizing photons s$^{-1}$ in Lyman continuum, equivalent to a single star type O \citep{watson2008}. Considering a model of H II region in expansion the neutral material accumulates between the ionization front and the shock front \citep{deharveng2010}, the ionized gas is surrounded by a shell of dense, neutral material hosting PAHs, the main responsible of 8.0 $\mu$m emission in infrared wavelengths. Therefore ionized gas appears confined by the bubble shell exhibited by 8.0 $\mu$m emission in red color.

% Fig 07

\subsubsection*{Cold and warm dust}

In Figure \ref{multi_wave} it has been displayed the set of emission maps at 4.5, 5.8, 8.0 and 24 $\mu$m. These emissions, corresponding to warm dust, allows to identify clearly the expanding HII region around the bubble \object{N10}. The upper panels, maps at 4.5, 5.8 and 8.0 $\mu$m, reveals the arc-shaped boarder of \object{N10}. 

\begin{figure*}
\begin{center}
\includegraphics[scale=0.41,angle=0]{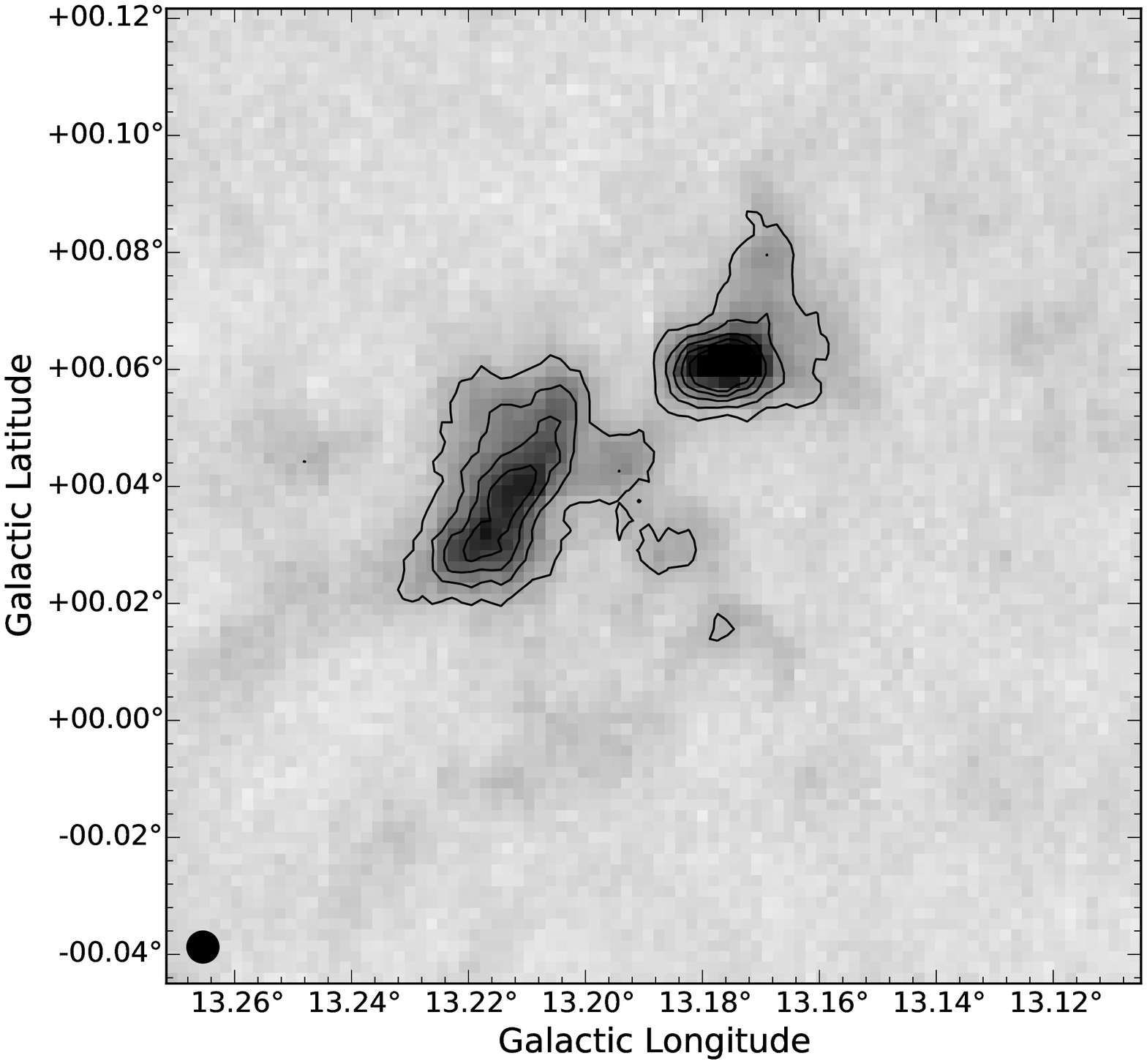}
\includegraphics[scale=0.41,angle=0]{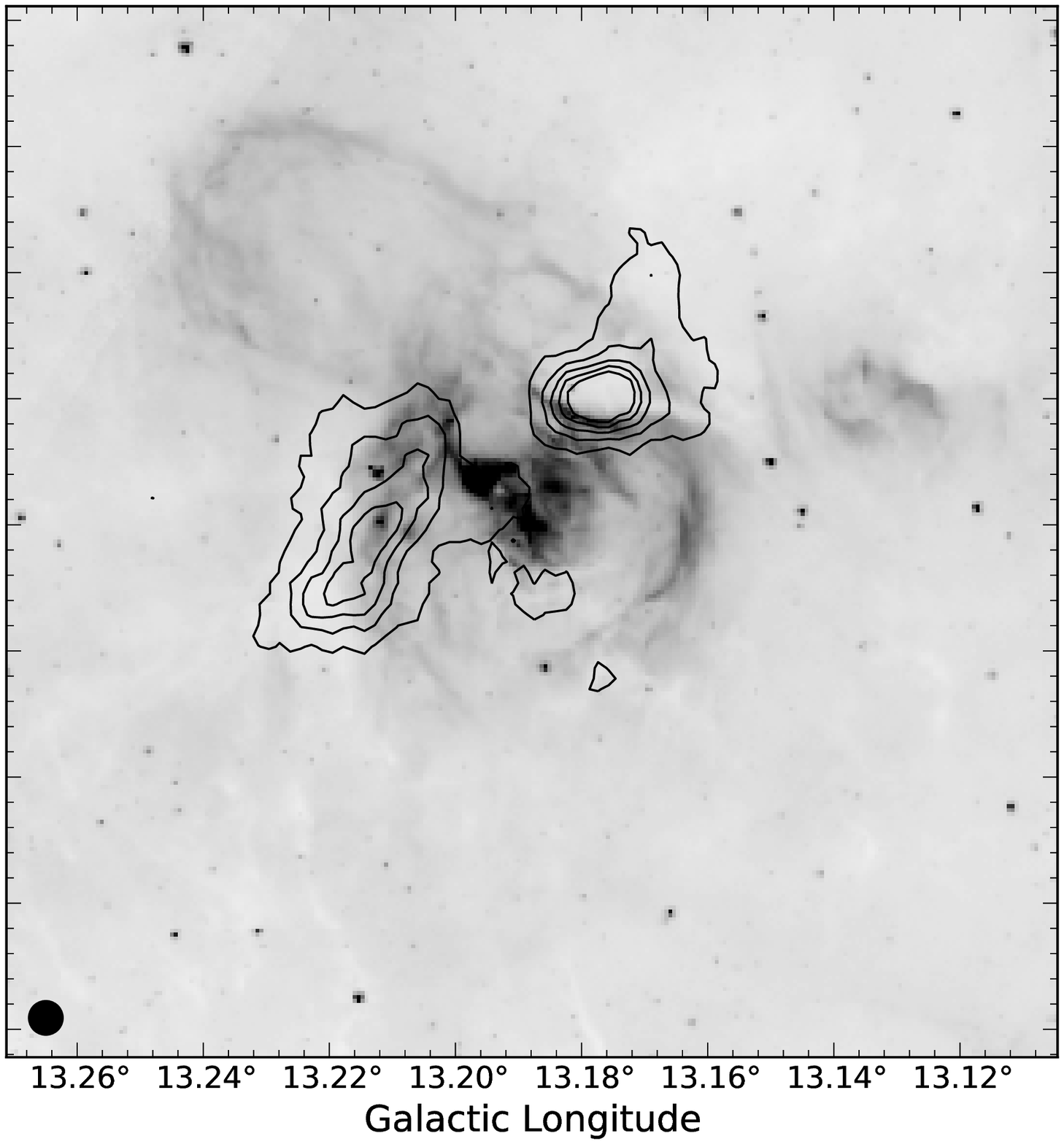}
\caption{\textit{Left panel}: LABOCA 870 $\mu$m emission showing the distribution of cold dust in grayscale, from 0 to 5 Jy beam$^{-1}$, with contours starting from 1 $\sigma$ to 5 $\sigma$, in steps of 1 $\sigma$ (1 $\sigma$ = 0.5 Jy beam$^{-1}$). \textit{Right panel}: Spitzer 8.0 $\mu$m image in background, overlaid with the same 870 $\mu$m contours of left figure. In both figures the black dot represents the beam size.}
\label{map8_cont870}
\end{center}
\end{figure*}

The thermal emission from cold dust is responsible for mainly continuum 870 $\mu$m distribution towards \object{N10}, while the emission at 8.0 $\mu$m originates by polyciclic aromatic hydrocarbons (PAHs) excited by UV photons. Figure \ref{map8_cont870} displays the emission at 870 $\mu$m in grayscale and contours for \object{N10} (left panel), and the same contours of the cold dust emission are superimposed on an Spitzer 8.0 $\mu$m image (right panel).  

The two 870 $\mu$m clumps coincide with $^{13}$CO molecular condensations detected at the peak velocity of 52 km s$^{-1}$ shown in upper right panel Figure \ref{8micron_co}. According \citet{dewangan2015a} the coincidence of distribution of the molecular gas, PAH and cold dust emission are an evidence of star-forming material around the bubble. We calculated the physical parameters for the denser condensation in the right of the Figure \ref{map8_cont870}. We here consider that the 870 $\mu$m radiation originates from the thermal radiation from dust grains. The total mass (dust and gas) in grams can be estimated following relation by \citet{hildebrand1983}:

\begin{eqnarray} \label{total-mass}
M_{tot} = 
100
\frac{\mathrm S_{870 \mu m} \ D^{2}}{\mathrm \kappa_{870 \mu m} \  B_{\mathrm 870 \mu m}(\mathrm T_{dust})},
\end{eqnarray}

\noindent where ${\mathrm S_{870 \mu m}}$ is the flux density of 870 $\mu$m emission in Jy, ${\mathrm \kappa_{870 \mu m}}$ is the dust opacity per unity mass at 870 $\mu$m and $B_{870 \mu m}$ is the Planck function for a given dust temperature ${\mathrm T_{dust}}$ in Jy. We assumed $\mathrm {T_{dust} = 20}$ K and ${\mathrm \kappa_{870 \mu m} = 1.8}$ cm$^{2}$ g$^{-1}$. We assumed a gas-to-dust ratio of 100 \citep{mathis1977, draine1985}.

We can calculate the $H_{2}$ column density $N(H_{2})$ using the following formula, by \cite{deharveng2010}:

\begin{eqnarray} \label{colunm-density}
\mathrm N(H_{2})\ = 
\frac{\mathrm 100 \ F_{870 \mu m}}{\mathrm \kappa_{870 \mu m} \ B_{\mathrm 870 \mu m}(\mathrm T_{dust}) \ 2.8 \ m_{H} \ \Omega_{beam}},
\end{eqnarray}

\noindent where ${\mathrm N(H_{2})}$ is in cm$^{-2}$, the surface brightness ${\mathrm F_{870 \mu m}}$ is in Jy beam$^{-1}$, the beam solid angle ${\mathrm \Omega_{beam}}$ is in steradians and the hydrogen mass m$_{H}$ is in grams.

We calculate the effective radius of the clumps as:

\begin{eqnarray} \label{radius}
{\mathrm R_{D}}\ = 
\Bigg( \frac{\mathrm \theta_{maj,D}}{2}  \times \frac{\mathrm \theta_{min,D}}{2} \Bigg)^{0.5},
\end{eqnarray}

\noindent where the major and minor deconvolved FWHM of the condensation (${\mathrm \theta_{maj,D}}$ and ${\mathrm \theta_{min,D}}$) was calculated as:

\begin{eqnarray} \label{deconvolved}
{\mathrm \theta_{maj,D}=\sqrt{\mathrm \theta_{maj}^{2} - \theta_{HPBW}^{2}}} , \quad 
{\mathrm \theta_{min,D}=\sqrt{\mathrm \theta_{min}^{2} - \theta_{HPBW}^{2}}}
\end{eqnarray}

\noindent where $\theta_{maj}$ and $\theta_{min}$ are the major and the minor FWHM sizes, respectively, was obtained by using the GreG/GILDAS fitting. We found that $\theta_{maj} = 44$ arcsec and $\theta_{min} = 32$ arcsec. The half-power beamwidth for $870 \ \mu m$ is $\theta_{HPBW} = 19.2$ arcsec. Therefore, $R_{min,D}$ = 25.6 arcsec and $R_{maj,D}$ = 39.6 arcsec (0.6 pc and 0.9 pc, respectively, at a distance of 4.7 kpc). According to Equation \ref{radius} the mean radius deconvolved is $R_{D} = 0.36$ pc, at a distance 4.7 kpc.

The average volume density for this condensation, assuming a spherical geometry, was calculated according \citet{duronea2015}:

\begin{eqnarray} \label{density-volume}
n(H_{2}) = \frac{\mathrm M_{tot}}{\mathrm 4/3 \ \pi \ R^{3}_{D} \ \mu \ m_{H}},
\end{eqnarray}

\noindent where $\mu$ is the mean molecular weight and $m_{H}$ is the mass of hydrogen atom. We assumed ${\mathrm \mu = 2.33}$ g and ${\mathrm m_{H} = 1.67 \ \times 10^{-24}}$ g.

In short, we found a column density $N(H_{2}) = 6.3 \times 10^{22}$ cm$^{-2}$, a total mass ${\mathrm M_{tot} = 240}$ M$_{\odot}$, the mean radius deconvolved is ${\mathrm R_{D} = 0.36}$ pc and the average volume density is ${\mathrm n(H_{2}) = 9.4 \times \ 10^{4}}$ cm$^{-3}$.

% Fig 08

\begin{figure*} 
\begin{center}
\includegraphics[scale=0.42,angle=0]{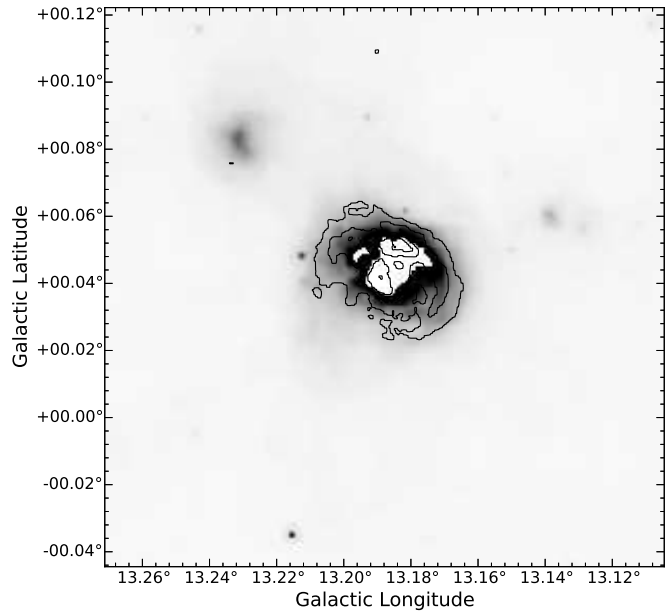}
\caption{24 $\mu$m distribution in grayscale from 0 to 2280 MJy sr$^{-1}$ with the same 20 cm contours of left panel in Figure \ref{map8_cont20}.}
\label{map24_cont20}
\end{center}
\end{figure*}

Warm dust can also be traced in this region by distribution of 24 $\mu$m emission in grayscale. VLA 20 cm emission is shown in contours. Warm dust and ionized emissions appears to be quite correlated as Figure \ref{map24_cont20} displays, as expected for HII regions \citep[e.g.][]{paladini2012}. 

We can notice that the IRAC 24 $\mu$m emission appears saturated close to the VLA 20 cm emission. This wavelength band concerns dust and gas ionized emission, which is shown in our results too. From Figure \ref{multi_wave} we can see that the saturated region is that both the IRAC and 20cm emissions are strong.  

% Fig 09

In resume, the map at 8.0 $\mu$m is very useful, since it traces the expanding arc in the upper left part of \object{N10}, this region is associated with ionized gas, excited PAHs and warm dust. 

On the other hand, the emission map at 20 cm exposes the central part of the HII region that is surrounded by the \object{N10} boarder; as would be expected, this region contains dust warmer than the expanding arcs. Furthermore, Figure \ref{map8_cont20} shows clearly the central part of the HII region.

\subsection{Distance}
\label{Distance}

Adopting a distance of 4.6 kpc from \citet{pandian2008} \citep[see also][]{deharveng2010}, the physical size of the ring is about 4.7 $\times$ 2.5 pc. Using this distance, \object{N10} is found to be close to the near extremity of the Galactic bar, a region of intensive star formation \citep[see e.g. the maps of the Galactic arms structure by][]{hou2014}. 

For this work, using circular Galactic rotation models \citep[e.g.][]{brand1993} is possible to compute near and far kinematic distances of the source; we have analyzed the kinematic distance 
ambiguity (KDA) and results shows the near kinematic one may be more reasonable.  

\cite{churchwell2006} argued that infrared bubbles are more likely located at their near kinematic distances, since objects on the far side of the Galactic disk would be obscured by interstellar extinction and contamination of other structures. 

Based on our CO observations and using the velocity of 52.6 km s$^{-1}$ (see Subsection \ref{obs-deriv-param}), we obtained near and far kinematic distances of 4.7 kpc and 11.3 kpc, respectively. This value is compatible with the near distance estimated by \cite{szymczak2000} d = 4.4 kpc, using the methanol maser emission (see Table \ref{velocities}). Since \object{N10} is located in the inner disk of the Galaxy, we adopt a 10\% uncertainty for the kinematic distance \citep{jinghua2014}, resulting in the value of 4.7$\pm$0.5 kpc for \object{N10}.

\subsection{Observed and derived parameters}
\label{obs-deriv-param}

\begin{figure*}
\begin{center}
\small
% \setcaptionmargin{1cm}
\includegraphics[scale=0.41]{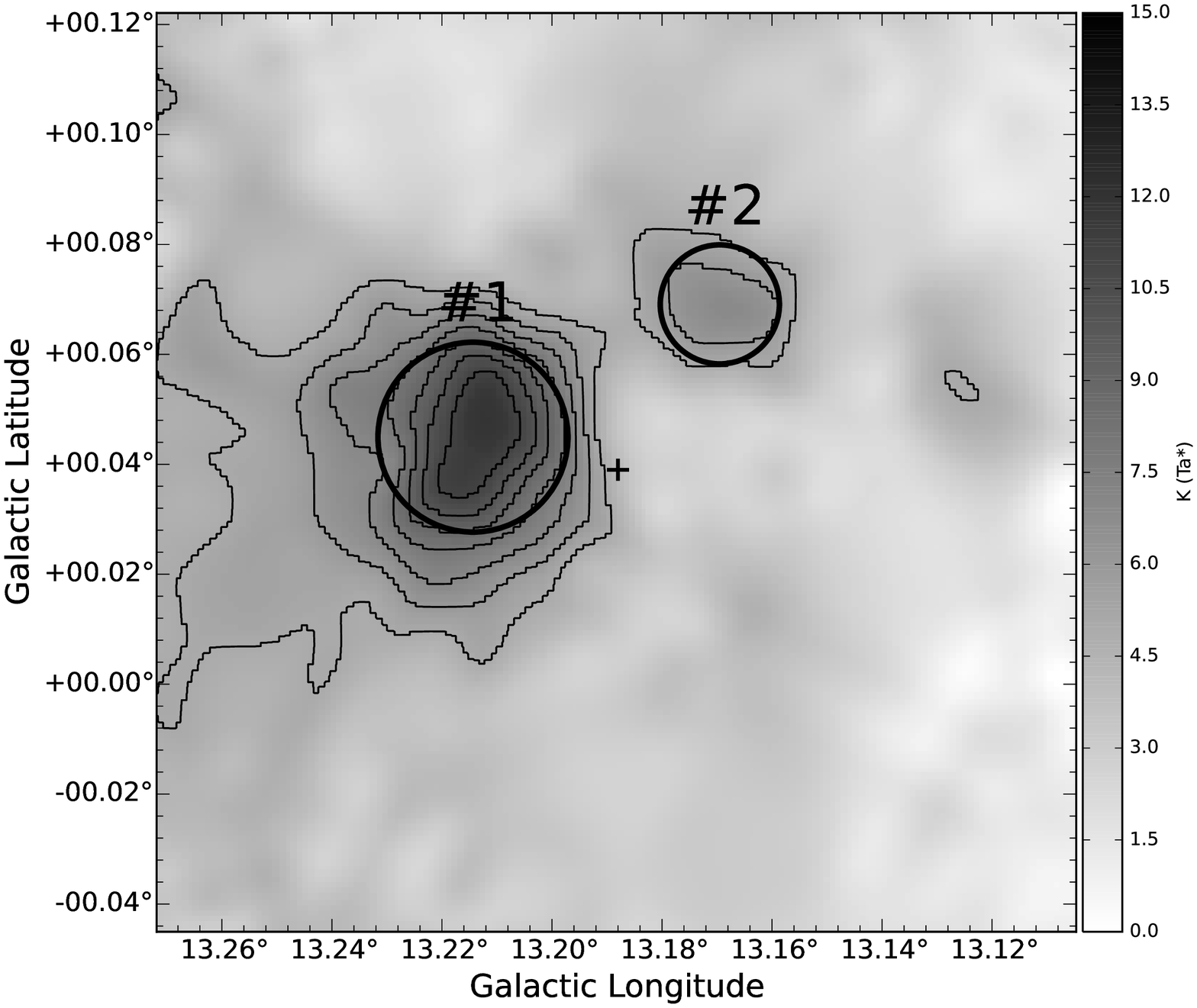}
\caption {Region of the clumps identified in $^{13}$CO intensity centered at 52 km s$^{-1}$, in grayscale. Clump \#1 and Clump \#2 seem to be physically associated with bubble N10. Black contours show also $^{13}$CO emission from 5 to 11-K, in steps of 1-K. The black cross indicates the center of the HII region inside N10.} 
\label{clumps}
\end{center}
\end{figure*}

In the following discussion we adopt the labels given in Figure \ref{clumps} to refer to identified condensations. Due to the poor spatial resolution of PMO (the beam size is $\sim$ 0.9 pc in our observations), the sizes of these molecular condensations could be smaller than 0.9 pc.

We used CLASS to calculate the parameters performing Gaussian fits to the average spectra obtained for the whole observed region. We obtained the centroid velocity (V$_{LSR}$), the antenna temperature (T$_{A}^{*}$), and the full width at half-maximum ($\Delta$V$_{FWHM}$). These observed parameters are shown in Table \ref{observed}.

% Fig 10

In order to understand the evolutionary status of molecular clumps, we derived the physical properties for the two clumps we identified in our CO observations. The mass of the clumps were estimated by using the Miriad software package \citep{sault1995}.

The mass of the molecular gas in a clump is calculated from the intensity of the emission line, under the local thermal equilibrium (LTE) assumption. With the radiation transfer equation \citet{garden1991} we derived the mass of the clumps under LTE condition, as follow:

\begin{eqnarray} 
\label{mlte}
M_{LTE} = \frac{4}{3} \ \pi  \ R^{3} \ m_{H_{2}} \ \mu_{g} \ n_{H_{2}}
\end{eqnarray}

\noindent where $M_{LTE}$ is given in solar masses $M_{\odot}$, $m_{H_{2}}$ is the mass of hydrogen molecule, $\mu_{g}$ = 1.36 is the mean atomic weight of the gas and ${n_{H_{2}} = N/2R}$ is the volume density. The mean radius of the clump is obtained by the relation $R = \sqrt[]{b_{maj} \times b_{min}}/2$, where $b_{maj}$ and $b_{min}$ are the sizes of the minor and major axes of the ellipse, respectively, obtained using the Miriad software.

We first estimated the value of the column density ${H_{2}}$ through the CO column density. According to \citet{garden1991} the column density $ N $ of a rigid, asymmetric linear molecule, under LTE condition, can be expressed by:

\begin{eqnarray} 
\label{N}
N &=& \frac{3 k}{8 \pi^{3}B\mu^{2}} \times \frac{exp[h \ B \ J(J+1) / k \ T_{ex}]}{(J+1)} \nonumber \\
& \times &   \frac{(T_{ex}+h B/3 k)}{[1 - exp(- h \nu/k T_{ex}]} \times \int \! \tau \, \mathrm{d}\nu
\end{eqnarray}

\noindent where B is the rotational constant, J is the rotational quantum number of the lower state of the observed transition and $\mu$ is the electric dipole of the molecule. $T_{ex}$ is the excitation temperature and $\tau$ is the optical depth from 48 to 53 km s$^{-1}$. 

Since the excitation temperature ($T_{ex}$) is measured as a function of brightness temperature ($T_{r}$), we estimate \citep{garden1991}:

\begin{eqnarray}
\label{Tr}
T_{r} &=& \frac{h\nu}{k} \times \Bigg[\frac{1}{exp(h\nu/kT_{ex})-1} - \frac{1}{exp(h\nu/kT_{bg})-1}\Bigg] \nonumber \\
& \times &  [1-exp(-\tau)]f
\end{eqnarray}

\noindent where $T_{r}$ is the brightness temperature and the temperature of the cosmic background radiation $T_{bg}$ = 2.73 K. Here we assume a filling factor of $f$ = 1.

Assuming that $T_{ex}$ is the same for $^{12}$CO and for $^{13}$CO, the optical depth for both lines can be obtained directly comparing the measure of its brightness temperatures $T_{r}$ \citep{garden1991}:

\begin{eqnarray} 
\label{tau}
\frac{T_{r}(^{12}CO)}{T_{r}(^{13}CO)} \approx \frac{1-exp(-\tau_{12})}{1-exp(-\tau_{13})}
\end{eqnarray}

We adopted an isotope ratio of [$^{12}$CO]/[$^{13}$CO] = 60 \citep{deharveng2008,wilson1994}, implying $\tau_{12}/\tau_{13} = 60$, and the canonical [CO]/[H$_{2}$] abundance ratio of 10$^{-4}$.

Thus we can estimate the optical depth from Equation \ref{tau}, and then, using it in Equation \ref{Tr}, and knowing the brightness temperature, we can estimate $T_{ex}$.

Using the estimated value of $T_{ex}$ for the line $^{13}$CO, we can finally obtain the hydrogen column density ($N_{H_{2}}$) using the equation \ref{N}. Thus, the intensity of the $^{13}$CO line traces the column density of the clumps \#1 and \#2, as listed in Table \ref{derived}.

% Table 03
\begin{deluxetable}{c c c c}
%\tabletypesize{\scriptsize}
\tablewidth{0pt}
\tablecolumns{4}  
\tablecaption{Observed line parameters obtained over the integrated area described in Figure \ref{spectra}, where V$_{lsr}$ is the centroid velocity of the main peak, the T$_{A}^{*}$ is the antenna temperature and $\Delta$V$_{FWHM}$ is the full width at half-maximum of lines.\label{observed}} 
\tablehead{
\colhead{} $\qquad$ & \colhead{V$_{lsr}$ } $\qquad$ & \colhead{T$_{A}^{*}$} $\qquad$ & \colhead{$\Delta$V$_{FWHM}$ } \\
\colhead{} $\qquad$ & \colhead{(km $s^{-1}$) $\qquad$} & \colhead{(K) } $\qquad$ & \colhead{(km $s^{-1}$)}
}
\startdata 
$^{12}$CO  $\qquad$ & 52.6  $\qquad$ & 8.3  $\qquad$ & 9.2 \\
$^{13}$CO  $\qquad$ & 52.6  $\qquad$ & 2.9  $\qquad$ & 6.1 \\
\enddata
\end{deluxetable}

% Table 04
\begin{deluxetable}{c c c c c c c c} 
%\tabletypesize{\scriptsize}
\tablecolumns{8} 
\tablewidth{0pt} 
\tablecaption{Derived parameters of the clumps observed in $^{13}$CO.\label{derived}} 
\tablehead{ 
\colhead{clump} & \colhead{R} & \colhead{T$_{ex}$} & \colhead{N$_{H_{2}}$} & \colhead{n$_{H_{2}}$} & \colhead{M$_{LTE}$} & \colhead{M$_{virial}$} & \colhead{M$_{Jeans}$}  \\
\colhead{} & \colhead{(pc)} & \colhead{(K)} & \colhead{(10$^{22}$ cm$^{-2}$)} & \colhead{(10$^{3}$ cm$^{-3}$)} & \colhead{(10$^{3}$ M$_{\odot}$)} & \colhead{(10$^{3}$ M$_{\odot}$)} & \colhead{(10$^{3}$ M$_{\odot}$)}
}
\startdata 
 \#1 & 1.1 & 16.8 & 4.1 & 4.2 & 2.6 & 9.3 & 7.0  \\ 
 \#2 & 1.2 & 12.9 & 3.3 & 3.9 & 1.5 & 7.8 & 4.7  \\
\enddata
\end{deluxetable}

% \tablenotemark{key letter(s)}
% \tablenotetext{alpha key}{text}

Obtaining the velocity dispersion and the mass under LTE assumption we can estimate the virial condition, by comparing the gas mass ($M_{LTE}$) with the virial mass ($M_{viral}$). In a cloud in which the temporal average kinetic energy is equal to half of the temporal average of the potential energy, the system is considered in virial equilibrium. The assumption that a gravitationally bound system is in virial equilibrium is widely used in astrophysics to estimate its mass \citep{huang1954}. The virial mass $ M_ {viral} $ is given by \citet{ungerechts2000} by the expression:

\begin{eqnarray} 
\label{m-vir}
\frac{M_{virial}}{M_{\odot}}= 2.10 \times 10^2 \Bigg( \frac{R}{pc} \Bigg) \Bigg( \frac{\Delta V_{FWHM}}{km s^{-1}} \Bigg)^2 
\end{eqnarray}

\noindent where R is the mean radius of the clump and $\Delta V_{FWHM}$ is the line width of $^{13}$CO line.

The Jeans mass $M_{J}$ is the mass is the mass above which a gas cloud will collapse, for a given density and temperature, when the gravitational attraction overcomes the pressure of the gas. It can be calculated according to \cite{stahler2005}:

\begin{eqnarray} \label{mjeans}
\frac{M_{Jeans}}{M_{\odot}} = \Bigg(\frac{T}{10 K} \Bigg)^{3/2} \Bigg( \frac{n_{H_{2}}}{10^{4} \ cm^{-3}} \Bigg)^{-1/2}
\end{eqnarray}

The results are presented in Table \ref{derived} where the columns 2 -- 7 list the following clump parameters: mean radius ($R$), excitation temperature ($T_{ex}$), column density ($N_{H_{2}}$), volume density ($n_{H_{2}}$), gas mass calculated under LTE assumption ($M_{LTE}$), virial mass ($M_{virial}$) and Jeans mass ($M_{Jeans}$). We discuss the clump status in Subsection \ref{surrounding-gas}.

\subsection{Identification of YSOs in the field of N10}
\label{Identification of YSOs in the field of N10}

The distribution of Young Stellar Objects (YSOs) plays a major role in the interpretation of the dynamics of star forming region. To identify the YSOs present in the field of \object{N10}, we adopted the method described by \citet{koenigleisawitz2014} (hereafter KL), based on the data of the Wide-field Infrared Survey Explorer \citep[WISE; see][]{wright2010}. In particular we used the AllWISE release \citep[Cutri et al. 2011\footnote{\citet{cutri2011}},][]{cutri2013}, which combined the data from the cryogenic and post-cryogenic phases of the survey, resulting in a catalog with enhanced sensitivity.  

The catalog was accessed through the VIZIER facility of the Strasbourg Data Center. The catalog contains infrared photometric data at 3.6, 4.9, 5.8 and 22 $\mu$m wavelengths, hereafter designated as w1, w2, w3 and w4 bands, respectively. In a first step, we selected all the objects situated in the area around \object{N10} that we explored, in the range of Galactic coordinates ${\mathrm 13.11^{\circ} < l < 13.27^{\circ}}$ and ${\mathrm -0.04^{\circ} < b < 0.12^{\circ}}$.

We found 565 WISE sources in this area. We next filtered this list of sources by applying a serie of quality criteria defined by KL, that they call the uncertainty/signal-to-noise/chi-squared criteria. The purpose of this is to avoid regions in the space of these parameters with relatively high probability of spurious catalog entry. Accordingly, the Class I YSOs are classified as such if their color matches with all the following criteria:

\begin{center}
\begin{itemize}
\item[] $w1 - w3 > 2.0$; 
\item[] $w1 - w2 > - 0.42 \times (w2 - w3) + 2.2$;
\item[] $w1 - w2 > 0.46 \times (w1 - w3) - 0.9$;
\item[] $w2 - w3 < 4.5$.
\end{itemize}
\end{center}

\noindent These conditions reflect the divisions in the SED slope ${\mathrm \alpha = d \ log(\lambda F_{\lambda})/d \ log \lambda}$.

The Class II objects were also classified according to KL, whose criteria are:

\begin{itemize}
\item[] $w1 - w2 > 0.25$;
\item[] $w1 - w2 < - 0.9 \times (w2 - w3) + 0.25$;
\item[] $w1 - w2 > 0.46 \times (w2 - w3) - 0.9$;
\item[] $w2 - w3 < 4.5$.
\end{itemize}

For w3 we kept only the condition of S/N larger than 5. It is considered that if a source satisfies the criteria of being a true source in any one of the bands, it has little probability of being a fake one and will be included in the final list.  After this filtering the list of sources reduced to 407 entries.

Next step was to separate the sources into Class I, Class II, Transition Disks and remaining objects. Following KL, the Class I and Class II YSOs are classified as such, based on the w1-w2 versus w2-w3 color-color diagram only. The regions of the diagram that are used to classify the  YSOs are defined by a number of frontier lines, shown in Figure \ref{ccdiag_01}. The equations of the lines are given by KL (their equations 12 to 20). We found 12 Class I stars and 91 Class II sources. The Transition Disk stars are selected separately by means of the  w1-w2 versus w2-w3 color-color diagram as shown in Figure \ref{ccdiag_02}. We found 131 transition disk stars. Note that the selection criteria follow an order of priority: a Class I object will remain Class I even if it also satisfies the criterion for Class II, and next the Class II selection prevails over the following selection. This is why we find many Class II objects in the box defining transition disk stars in Figure \ref{ccdiag_02}: they were classified Class II in the previous step, on the basis of the different color-color plot.  

This selection of Class I objects is robust, since these objects are well separated from the other classes in the color diagrams. Furthermore, we made experiments with another classification scheme available in the literature \citep[using Spitzer data, e.g.][]{gutermuth2009} and the same Class I objects were retrieved. On the other hand, the distinction between Class II and Transition Disk is a little arbitrary, as we can see some overlap in Figures \ref{ccdiag_01} and \ref{ccdiag_02}. In the samples of objects previously known to belong to given classes, used by KL to decide the position of the frontier lines in Figures \ref{ccdiag_01} and \ref{ccdiag_02}, one can see a number of transition disk sources in the locus of Class II and vice versa. So, one must consider that the decision to attribute sources to one or the other classes is only valid in a statistical sense, being correct in about 70\% of the cases.

\subsection{SED fitting}

We have compared the position of Class I YSOs, the most embedded young stellar sources, with the molecular distribution and the objects identified from \#1 to \#9 (see Table \ref{all_ysos}) are more likely to be physically related to N10 molecular structure. We have fitted their spectral energy distribution (SED) by using the online tool developed by \cite{robitaille2007}. Radiation transfer models were fitted to observational data extracted from WISE catalog based on a $\chi^{2}$ test. We selected models for which $\mathrm \chi^{2} - \chi^{2}_{min} < 3 n$, where $\mathrm \chi^{2}_{min}$ is the minimum value and $n$ is the number of input data. 

\begin{figure*}
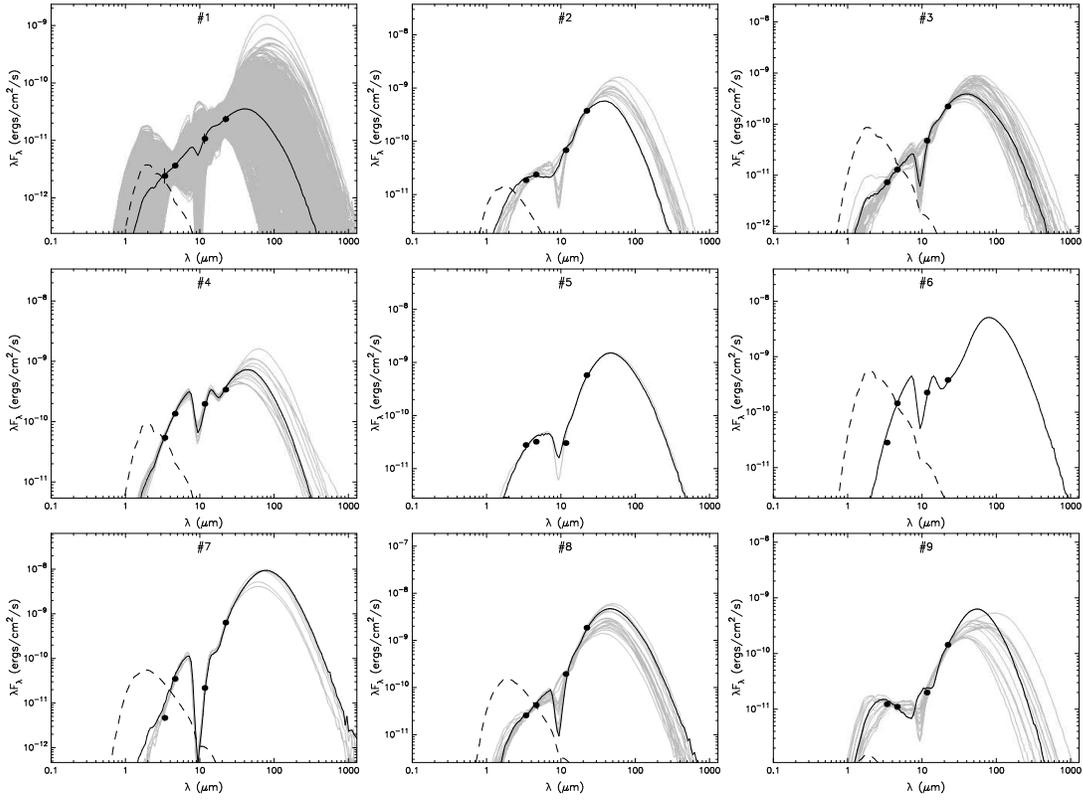

\begin{center}
% \setcaptionmargin{1cm}
\includegraphics[scale=0.4,angle=0]{classI_01.eps}
\includegraphics[scale=0.4,angle=0]{classI_02.eps}
\includegraphics[scale=0.4,angle=0]{classI_03.eps}
\includegraphics[scale=0.4,angle=0]{classI_04.eps}
\includegraphics[scale=0.4,angle=0]{classI_05.eps}
\includegraphics[scale=0.4,angle=0]{classI_06.eps}
\includegraphics[scale=0.4,angle=0]{classI_07.eps}
\includegraphics[scale=0.4,angle=0]{classI_08.eps}
\includegraphics[scale=0.4,angle=0]{classI_09.eps}
\caption{The filled circles show the input fluxes. The black line shows the best fit. The dashed line shows the stellar photosphere corresponding to the central source of the best fitting model, as it would loo in the absence of circunstellar dust (but including interstellar extinction).} 
\label{sed_fitting}
\end{center}
\end{figure*}

\begin{figure*}[ht]
\begin{center}
% \setcaptionmargin{1cm}
\includegraphics[scale=0.55,angle=0]{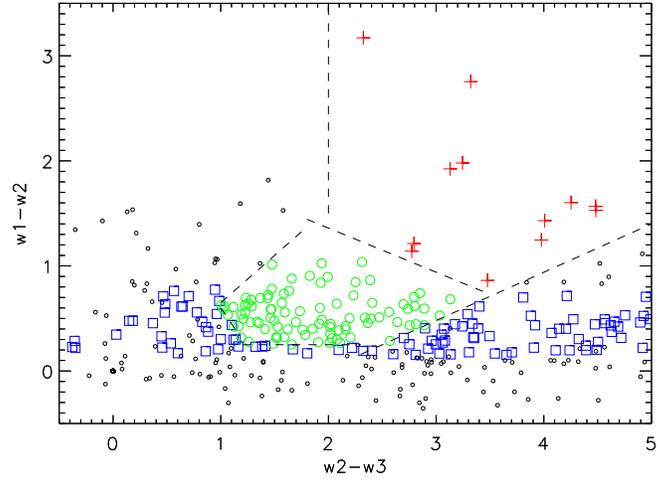}
\caption {Color-color diagram w1-w2 versus w2-w3 of the candidate YSOs around N10. Red crosses: Class I objects; green circles: Class II objects, blue squares: transition disk objects. The dashed lines indicate the limits of the regions according to HL. The transition disks were not defined by means of this diagram. The remaining objects are indicated by black dots.} 
\label{ccdiag_01}
\end{center}
\end{figure*}

\begin{figure*}[ht]
\begin{center}
% \setcaptionmargin{1cm}
\includegraphics[scale=0.55,angle=0]{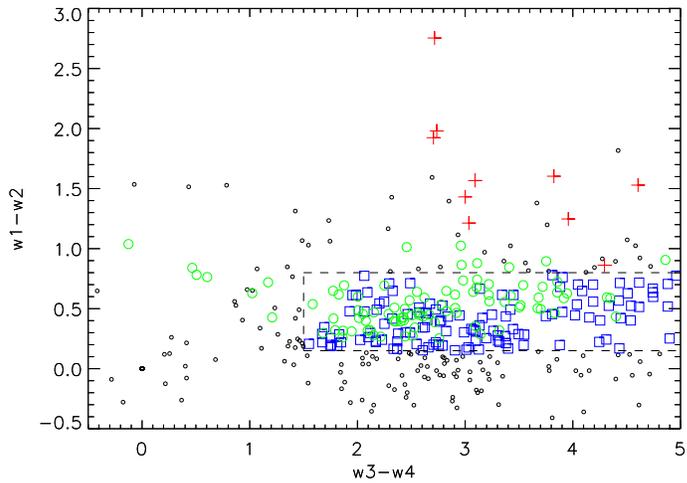}
\caption {Color-color diagram w1-w2 versus w3-w4, like Figure \ref{ccdiag_01}. The box used to define transition-disk objects (in blue) also contains many Class II objects (green circles) because these objects were defined to be Class II in a previous step using another color-color diagram (Figure \ref{ccdiag_01}).} 
\label{ccdiag_02}
\end{center}
\end{figure*}

The fitting was performed using fluxes from WISE data, distance ranges from 4.23 to 5.17 kpc. Interstellar extinction in the direction of \object{N10} was predict to be approximately 10.7 mag according to model S of \cite{amores2005}, values adopted were from 9.7 to 11.7 mag. The best fit are shown in Figure \ref{sed_fitting}. Resulting values for model parameters are given in Table \ref{model-parameters}. We found that Class I YSOs have stellar mass ranging from $\sim$1 to $\sim$13 M$_{\odot}$, stellar temperature $\sim 4000$ -- $20000$ K, total luminosity $\sim 3 \times 10^{1}$ -- $1 \times 10^{3}$ L$_{\odot}$, envelope accretion rate $\sim 9 \times 10^{-8}$ -- $3 \times 10^{-3}$ M$_{\odot}$ yr$^{-1}$, disk mass $\sim 7 \times 10^{-3}$ -- $6 \times 10^{-1}$ M$_{\odot}$ and stellar ages from $\sim 2\times 10^{3}$ to $\sim 1\times 10^{6}$ yr.

% Fig 11

% Table 05
\begin{deluxetable}{lccccccccc}
\tabletypesize{\small}
\rotate
\tablecaption{Physical parameters derived from \citet{robitaille2007} model, for associated Class I objects candidates.\label{model-parameters}}
\tablewidth{0pt}
\tablehead{
\colhead{} & \multicolumn{9}{c}{Class I YSOs $^{a}$} \\
\colhead{Parameters} & \colhead{1} & \colhead{2} & \colhead{3} & \colhead{4} & \colhead{5} & \colhead{6} & \colhead{7} & \colhead{8} & \colhead{9}
}
\startdata
Stellar Mass (M$_{\odot}$)    & 1.42    & 6.53    & 7.07    & 6.08    & 7.42    & 9.86    & 12.94     & 12.89   & 6.43  \\

Stellar Temperature (K)	    & 4173	  & 10107	& 4293	  & 4148	& 20167	  & 4260	& 12262	    & 7471	  & 19400 \\

Total Luminosity  (L$_{\odot}$) $^{b}$ & 3.11$\times10^{1}$ & 1.02$\times10^{3}$ & 5.45$\times10^{2}$ & 5.96$\times10^{2}$ & 2.33$\times10^{3}$ & 2.47$\times10^{3}$ & 1.13$\times10^{4}$ & 7.31$\times10^{3}$ & 1.19$\times10^{3}$  \\

$\dot{M}_{env}$ (M$_{\odot}$ yr$^{-1}$) & 2.49$\times10^{-5}$	& 1.20$\times10^{-5}$	& 4.70$\times10^{-5}$	& 1.19$\times10^{-5}$	& 7.60$\times10^{-5}$	& 1.12$\times10^{-5}$	& 3.61$\times10^{-3}$	& 2.25$\times10^{-4}$	& 8.62$\times10^{-8}$ \\

Disk Mass (M$_{\odot}$) $^{c}$ & 1.01$\times10^{-2}$ & 3.05$\times10^{-2}$ & 3.38$\times10^{-2}$ & 4.45$\times10^{-3}$ & 1.46$\times10^{-3}$ & 7.71$\times10^{-3}$ & 1.49$\times10^{-1}$ & 6.09$\times10^{-1}$ & 5.25$\times10^{-5}$ \\

Stellar Age (yr)            & 1.24$\times10^{4}$ & 2.77$\times10^{5}$ & 7.07$\times10^{3}$ & 1.84$\times10^{3}$ & 2.97$\times10^{5}$  & 2.41$\times10^{3}$ & 2.88$\times10^{4}$  & 1.60$\times10^{4}$  & 1.02$\times10^{6}$  \\
\enddata
\tablenotetext{a}{Identification from Table \ref{all_ysos}}
\tablenotetext{b}{Total system luminosity.}
\tablenotetext{c}{Envelope and ambient density mass, in solar masses.}
\end{deluxetable}

% Table 06 
\begin{deluxetable}{c c c c c c c c c}
\tabletypesize{\footnotesize}
% \rotate
\tablecaption{Young stellar objects (YSO) candidates towards the N10 region. \label{all_ysos}}
\tablewidth{0pt}
\tablehead{
\colhead{Candidate} & \colhead{l} & \colhead{b} & Identification & \multicolumn{4}{c}{Fluxes [mag]} & \colhead{Classification}\\
\#        & [$^{\circ}$] & [$^{\circ}$] &  & [3.4] & [4.6] & [12] & [22]    &  
}
\startdata

1   & 13.1141 & 0.0566  & J181354.14-173204.6 & 12.651 & 11.220 & 7.212  & 4.212  & Class I         \\
2   & 13.1385 & 0.0593  & J181356.48-173043.0 & 10.428 &  9.181 & 5.204  & 1.246  & Class I         \\
3   & 13.1383 & 0.0566  & J181357.08-173048.2 & 11.451 &  9.848 & 5.595  & 1.771  & Class I         \\
4   & 13.2124 & 0.0402  & J181409.60-172722.1 &  9.274 &  7.294 & 4.049  & 1.312  & Class I         \\ 
5   & 13.2007 & 0.0286  & J181410.77-172819.3 &  9.995 &  8.855 & 6.081  & 0.734  & Class I         \\
6   & 13.2126 & 0.0477  & J181407.98-172708.4 &  9.978 &  7.224 & 3.903  & 1.187  & Class I         \\
7   & 13.1818 & 0.0608  & J181401.37-172823.3 & 11.9\tablecomments{Table \ref{all_ysos} is published in its entirety in the electronic edition of the {\it Astrophysical Journal}.  A portion is shown here for guidance regarding its form and content.}37 &  8.766 & 6.442  & 0.629  & Class I         \\ 
8   & 13.2015 & 0.0554  & J181404.94-172730.5 & 10.081 &  8.552 & 4.069  &-0.537  & Class I         \\ 
9   & 13.2016 & 0.0681  & J181402.15-172708.2 & 10.877 & 10.015 & 6.538  & 2.244  & Class I         \\
10  & 13.2155 & -0.0350 & J181426.60-172921.9 &  9.291 &  7.367 & 4.237  & 1.531  & Class I         \\
11  & 13.2396 & -0.0057 & J181423.03-172715.2 & 12.263 & 11.050 & 8.256  & 5.221  & Class I         \\
12  & 13.1931 & 0.0888  & J181356.55-172659.5 & 12.308 & 10.741 & 6.260  & 3.168  & Class I         \\ 
    &         &         &                     &        &        &        &        &                 \\
13  & 13.1114 & 0.1134  & J181341.27-173035.2 & 10.397 & 9.934  & 8.657  & 4.825  & Class II	    \\
14  & 13.1135 & -0.0134 & J181409.56-173407.1 & 11.620 & 11.198 & 9.670  & 7.655  & Class II	    \\
15  & 13.1389 & -0.0279 & J181415.81-173311.7 & 11.511 & 11.061 & 9.146  & 6.480  & Class II	    \\
16  & 13.1384 & -0.0162 & J181413.16-173253.3 & 11.235 & 10.748 & 8.809  & 6.800  & Class II	    \\
17  & 13.1147 & -0.0195 & J181411.06-173413.8 & 11.343 & 11.046 & 8.968  & 7.287  & Class II	    \\
18  & 13.1449 & -0.0193 & J181414.65-173237.9 & 11.031 & 10.757 & 8.852  & 6.340  & Class II	    \\
19  & 13.1527 & -0.0295 & J181417.83-173231.1 & 10.529 & 10.021 & 7.684  & 5.341  & Class II	    \\
20  & 13.1543 & -0.0234 & J181416.66-173215.4 & 10.758 & 10.158 & 8.237  & 6.232  & Class II	    \\
21  & 13.1647 & -0.0310 & J181419.59-173155.6 & 11.511 & 10.498 & 9.022  & 6.565  & Class II	    \\
22  & 13.1577 & -0.0263 & J181417.72-173209.5 & 10.568 & 10.225 & 8.441  & 5.908  & Class II	    \\
23  & 13.1919 & -0.0358 & J181423.94-173037.7 & 10.433 & 10.010 & 8.891  & 6.481  & Class II	    \\
24  & 13.1704 & -0.0343 & J181421.01-173143.2 & 10.077 & 9.621  & 8.265  & 5.976  & Class II	    \\
25  & 13.2038 & -0.0395 & J181426.20-173006.5 & 8.941  & 8.625  & 7.104  & 5.237  & Class II	    \\
26  & 13.1103 & 0.0155  & J181402.78-173327.7 & 10.560 & 10.300 & 8.813  & 6.951  & Class II	    \\
27  & 13.1370 & -0.0034 & J181410.17-173235.5 & 9.904  & 9.313  & 8.291  & 5.669  & Class II	    \\
28  & 13.1452 & -0.0016 & J181410.76-173206.7 & 10.542 & 10.097 & 7.285  & 4.779  & Class II	    \\
29  & 13.1257 & -0.0028 & J181408.67-173310.3 & 11.147 & 10.754 & 8.995  & 6.637  & Class II	    \\
30  & 13.1362 & -0.0003 & J181409.38-173232.9 & 11.275 & 10.635 & 8.559  & 5.484  & Class II	    \\
31  & 13.1269 & 0.0325  & J181401.03-173205.8 & 10.224 & 9.903  & 8.315  & 6.522  & Class II	    \\
32  & 13.1505 & -0.0153 & J181414.43-173213.6 & 10.636 & 10.131 & 8.669  & 5.761  & Class II	    \\
33  & 13.1611 & -0.0034 & J181413.08-173119.4 & 10.832 & 10.136 & 8.904  & 5.365  & Class II	    \\
34  & 13.1541 & 0.0016  & J181411.12-173133.1 & 10.988 & 10.541 & 8.519  & 5.942  & Class II	    \\
35  & 13.1907 & -0.0196 & J181420.20-173013.7 & 10.467 & 9.685  & 8.212  & 7.706  & Class II	    \\
36  & 13.1771 & -0.0256 & J181419.91-173106.9 & 10.485 & 10.058 & 8.892  & 7.684  & Class II	    \\
37  & 13.1959 & -0.0330 & J181423.81-173020.5 & 10.700 & 10.302 & 8.892  & 6.814  & Class II	    \\
38  & 13.1979 & -0.0191 & J181420.96-172950.1 & 11.156 & 10.463 & 9.027  & 6.777  & Class II	    \\
39  & 13.1797 & -0.0231 & J181419.68-173054.4 & 10.657 & 10.390 & 8.281  & 6.066  & Class II	    \\
40  & 13.2025 & -0.0207 & J181421.88-172938.4 & 10.453 & 9.805  & 8.237  & 5.412  & Class II	    \\
41  & 13.1850 & -0.0311 & J181422.08-173051.7 & 11.027 & 10.458 & 8.722  & 6.238  & Class II	    \\
42  & 13.1763 & -0.0096 & J181416.28-173041.9 & 9.523  & 8.906  & 7.872  & 4.966  & Class II	    \\
43  & 13.1529 & 0.0133  & J181408.39-173116.8 & 11.277 & 10.554 & 9.114  & 5.257  & Class II	    \\
44  & 13.1778 & 0.0060  & J181413.01-173010.3 & 9.489  & 8.862  & 7.858  & 3.911  & Class II	    \\
45  & 13.1627 & 0.0214  & J181407.78-173031.9 & 9.323  & 8.688  & 5.991  & 2.136  & Class II	    \\
46  & 13.1732 & 0.0086  & J181411.88-173020.6 & 10.801 & 10.366 & 7.612  & 3.212  & Class II	    \\
47  & 13.1599 & 0.0238  & J181406.92-173036.4 & 9.450  & 8.858  & 6.121  & 2.438  & Class II	    \\
48  & 13.1904 & 0.0215  & J181411.09-172903.8 & 11.423 & 10.865 & 9.270  & 2.260  & Class II	    \\
49  & 13.1951 & 0.0128  & J181413.58-172903.9 & 9.707  & 8.801  & 6.974  & 2.114  & Class II	    \\
50  & 13.2076 & 0.0148  & J181414.64-172821.2 & 9.698  & 9.104  & 6.214  & 1.898  & Class II	    \\
51  & 13.1538 & 0.0598  & J181358.24-172953.7 & 9.997  & 9.382  & 8.095  & 4.876  & Class II	    \\
52  & 13.1348 & 0.0557  & J181356.85-173101.0 & 9.628  & 8.946  & 5.814  & 2.094  & Class II	    \\
53  & 13.1236 & 0.0950  & J181346.80-173028.4 & 10.652 & 10.249 & 8.394  & 5.963  & Class II	    \\
54  & 13.1157 & 0.0930  & J181346.30-173057.0 & 11.448 & 11.137 & 9.105  & 7.158  & Class II	    \\
55  & 13.1554 & 0.0962  & J181350.37-172845.8 & 10.915 & 10.286 & 9.060  & 8.037  & Class II	    \\
56  & 13.2163 & 0.0706  & J181403.36-172617.6 & 10.970 & 10.459 & 7.987  & 4.483  & Class II	    \\
57  & 13.2151 & 0.0670  & J181404.00-172627.5 & 11.001 & 10.383 & 7.583  & 5.756  & Class II	    \\
58  & 13.2071 & -0.0271 & J181423.84-172934.7 & 9.319  & 8.922  & 7.300  & 4.959  & Class II	    \\
59  & 13.2185 & -0.0141 & J181422.34-172836.2 & 10.337 & 9.983  & 7.966  & 4.871  & Class II	    \\
60  & 13.2078 & -0.0178 & J181421.88-172916.6 & 9.124  & 8.589  & 7.523  & 4.744  & Class II	    \\
61  & 13.2179 & -0.0213 & J181423.85-172850.7 & 10.761 & 10.112 & 7.705  & 5.931  & Class II	    \\
62  & 13.2204 & 0.0050  & J181418.36-172757.4 & 10.030 & 9.534  & 8.327  & 4.625  & Class II	    \\
63  & 13.2143 & 0.0017  & J181418.35-172822.5 & 10.872 & 10.285 & 8.820  & 4.903  & Class II	    \\
64  & 13.2536 & -0.0052 & J181424.60-172629.9 & 10.132 & 9.667  & 7.969  & 5.408  & Class II	    \\
65  & 13.2532 & 0.0007  & J181423.25-172621.3 & 11.055 & 10.557 & 8.166  & 5.727  & Class II	    \\
66  & 13.2621 & 0.0030  & J181423.79-172549.1 & 11.615 & 10.736 & 9.077  & 5.965  & Class II	    \\
67  & 13.2553 & -0.0207 & J181428.22-172651.4 & 9.615  & 9.146  & 7.810  & 5.657  & Class II	    \\
68  & 13.2630 & -0.0075 & J181426.23-172604.1 & 11.537 & 10.949 & 9.479  & 5.131  & Class II	    \\
69  & 13.2440 & -0.0054 & J181423.50-172700.7 & 10.056 & 9.031  & 7.095  & 4.135  & Class II	    \\
70  & 13.2507 & -0.0041 & J181424.00-172637.2 & 10.300 & 9.913  & 7.819  & 5.102  & Class II	    \\
71  & 13.2489 & 0.0039  & J181422.01-172629.3 & 7.988  & 7.294  & 5.927  & 4.050  & Class II	    \\
72  & 13.2484 & -0.0065 & J181424.27-172648.7 & 10.165 & 9.770  & 7.610  & 5.207  & Class II	    \\
73  & 13.2669 & -0.0343 & J181432.62-172638.0 & 10.417 & 9.654  & 8.019  & 7.414  & Class II	    \\
74  & 13.2583 & -0.0373 & J181432.25-172710.3 & 11.407 & 10.567 & 8.639  & 8.173  & Class II	    \\
75  & 13.2301 & 0.0140  & J181417.54-172711.3 & 10.361 & 9.642  & 7.410  & 6.239  & Class II	    \\
76  & 13.2255 & 0.0156  & J181416.61-172723.1 & 9.430  & 8.754  & 7.473  & 4.366  & Class II	    \\
77  & 13.2451 & 0.0341  & J181414.88-172549.3 & 10.995 & 10.254 & 7.405  & 4.293  & Class II	    \\
78  & 13.2404 & 0.0107  & J181419.49-172644.6 & 10.908 & 9.870  & 7.559  & 7.685  & Class II	    \\
79  & 13.2343 & 0.0419  & J181411.87-172610.1 & 9.853  & 9.426  & 7.266  & 4.838  & Class II	    \\
80  & 13.2499 & 0.0332  & J181415.66-172535.5 & 10.933 & 10.088 & 7.836  & 4.427  & Class II	    \\
81  & 13.2636 & 0.0175  & J181420.79-172519.2 & 9.982  & 9.342  & 8.087  & 5.534  & Class II	    \\
82  & 13.2574 & 0.0339  & J181416.42-172510.8 & 11.900 & 11.196 & 8.401  & 5.181  & Class II	    \\
83  & 13.2392 & 0.0430  & J181412.20-172552.7 & 10.673 & 10.118 & 8.891  & 5.524  & Class II	    \\
84  & 13.2558 & 0.0378  & J181415.36-172509.1 & 11.519 & 11.054 & 8.415  & 5.495  & Class II	    \\
85  & 13.1292 & 0.1156  & J181342.93-172935.3 & 10.796 & 10.527 & 8.517  & 6.073  & Class II	    \\
86  & 13.1461 & 0.1132  & J181345.50-172845.9 & 11.101 & 10.645 & 8.279  & 6.229  & Class II	    \\
87  & 13.1350 & 0.1103  & J181344.81-172926.0 & 8.488  & 7.950  & 6.759  & 5.174  & Class II	    \\
88  & 13.1616 & 0.1054  & J181349.09-172810.6 & 11.137 & 10.402 & 8.815  & 5.834  & Class II	    \\
89  & 13.1883 & 0.0892  & J181355.88-172714.1 & 7.892  & 7.571  & 6.055  & 3.938  & Class II	    \\
90  & 13.1815 & 0.1000  & J181352.66-172716.9 & 10.393 & 9.975  & 7.960  & 6.182  & Class II	    \\
91  & 13.1945 & 0.1155  & J181350.82-172609.0 & 10.328 & 10.050 & 8.788  & 6.638  & Class II	    \\
92  & 13.1833 & 0.1066  & J181351.44-172659.9 & 9.333  & 9.015  & 7.827  & 5.732  & Class II	    \\
93  & 13.2189 & 0.0964  & J181357.99-172524.8 & 9.412  & 8.517  & 6.523  & 2.770  & Class II	    \\
94  & 13.2282 & 0.0955  & J181359.30-172456.9 & 10.304 & 9.777  & 6.744  & 3.229  & Class II	    \\
95  & 13.2126 & 0.0974  & J181357.00-172543.1 & 9.804  & 9.485  & 7.280  & 3.962  & Class II	    \\
96  & 13.2128 & 0.1152  & J181353.08-172511.7 & 10.206 & 9.920  & 7.349  & 4.207  & Class II	    \\
97  & 13.2164 & 0.1159  & J181353.36-172459.1 & 9.300  & 10.050 & 8.821  & 4.499  & Class II	    \\
98  & 13.2433 & 0.0710  & J181406.52-172451.6 & 9.465  & 9.015  & 6.765  & 3.092  & Class II	    \\
99  & 13.2482 & 0.0971  & J181401.34-172351.1 & 9.776  & 9.168  & 8.050  & 4.627  & Class II	    \\
100 & 13.2679 & 0.0722  & J181409.21-172331.5 & 11.125 & 10.261 & 7.874  & 4.907  & Class II	    \\
101 & 13.2659 & 0.0771  & J181407.89-172329.7 & 9.496  & 9.243  & 7.583  & 4.824  & Class II	    \\
102 & 13.2420 & 0.1051  & J181358.82-172356.7 & 10.622 & 10.187 & 7.297  & 4.686  & Class II	    \\
103 & 13.2390 & 0.1185  & J181355.51-172343.3 & 10.482 & 9.921  & 7.358  & 4.132  & Class II	    \\
    &         &         &                     &        &        &        &        &                 \\
104 & 13.1380 & -0.0254 & J181415.16-173310.3 & 11.555 & 11.301 & 8.303  & 6.554  & Transition Disk \\
105 & 13.1131 & -0.0255 & J181412.18-173429.2 & 9.828  & 9.621  & 8.645  & 7.095  & Transition Disk \\
106 & 13.1302 & -0.0333 & J181415.95-173348.5 & 9.932  & 9.692  & 8.278  & 6.252  & Transition Disk \\
107 & 13.1319 & -0.0260 & J181414.56-173330.7 & 9.684  & 9.453  & 8.130  & 5.958  & Transition Disk \\
108 & 13.1630 & -0.0390 & J181421.17-173214.8 & 8.383  & 7.905  & 7.722  & 5.794  & Transition Disk \\
109 & 13.1434 & -0.0357 & J181418.08-173311.0 & 10.852 & 10.078 & 9.131  & 7.067  & Transition Disk \\
110 & 13.1559 & -0.0304 & J181418.41-173222.5 & 8.913  & 8.355  & 7.577  & 5.329  & Transition Disk \\
111 & 13.1591 & -0.0291 & J181418.52-173210.1 & 8.583  & 8.237  & 8.206  & 6.513  & Transition Disk \\
112 & 13.1685 & -0.0357 & J181421.10-173151.7 & 9.228  & 8.592  & 8.107  & 6.014  & Transition Disk \\
113 & 13.1178 & -0.0041 & J181408.01-173337.4 & 10.642 & 9.930  & 9.228  & 6.738  & Transition Disk \\
114 & 13.1414 & -0.0040 & J181410.84-173222.7 & 11.260 & 10.857 & 7.784  & 5.008  & Transition Disk \\
115 & 13.1429 & -0.0156 & J181413.57-173237.9 & 8.265  & 8.045  & 8.394  & 6.546  & Transition Disk \\
116 & 13.1471 & 0.0039  & J181409.76-173151.2 & 10.965 & 10.559 & 7.662  & 5.014  & Transition Disk \\
117 & 13.1338 & 0.0047  & J181408.00-173231.7 & 11.836 & 11.295 & 8.002  & 5.328  & Transition Disk \\
118 & 13.1148 & 0.0518  & J181355.31-173210.8 & 11.336 & 11.114 & 7.206  & 5.035  & Transition Disk \\
119 & 13.1511 & 0.0367  & J181403.02-173042.0 & 10.622 & 10.453 & 9.851  & 5.419  & Transition Disk \\
120 & 13.1361 & 0.0203  & J181404.83-173157.6 & 7.942  & 7.328  & 6.682  & 4.696  & Transition Disk \\
121 & 13.1737 & -0.0160 & J181417.38-173101.2 & 11.805 & 11.355 & 8.017  & 4.768  & Transition Disk \\
122 & 13.1491 & -0.0062 & J181412.24-173202.1 & 9.232  & 9.026  & 7.356  & 4.807  & Transition Disk \\
123 & 13.1801 & -0.0195 & J181418.90-173047.1 & 9.321  & 8.778  & 7.815  & 5.611  & Transition Disk \\
124 & 13.1889 & -0.0263 & J181421.47-173031.1 & 11.226 & 10.936 & 7.854  & 6.182  & Transition Disk \\
125 & 13.1775 & -0.0162 & J181417.88-173049.7 & 10.607 & 10.353 & 7.602  & 4.404  & Transition Disk \\
126 & 13.1810 & -0.0299 & J181421.31-173102.2 & 12.121 & 11.943 & 8.638  & 5.979  & Transition Disk \\
127 & 13.1864 & -0.0094 & J181417.43-173009.7 & 10.567 & 10.264 & 9.263  & 6.273  & Transition Disk \\
128 & 13.1724 & 0.0203  & J181409.19-173002.7 & 11.192 & 10.765 & 5.473  & 1.320  & Transition Disk \\
129 & 13.1668 & 0.0212  & J181408.31-173019.1 & 10.302 & 9.588  & 5.375  & 0.938  & Transition Disk \\
130 & 13.1777 & 0.0131  & J181411.42-172958.4 & 10.984 & 10.516 & 9.688  & 3.001  & Transition Disk \\
131 & 13.1698 & 0.0259  & J181407.64-173001.4 & 9.631  & 9.229  & 5.086  & 0.549  & Transition Disk \\
132 & 13.1665 & 0.0306  & J181406.20-173003.7 & 10.611 & 9.836  & 4.188  & -0.772 & Transition Disk \\
133 & 13.1861 & 0.0172  & J181411.53-172925.0 & 7.226  & 6.464  & 5.898  & 1.992  & Transition Disk \\
134 & 13.2099 & 0.0177  & J181414.29-172808.9 & 11.603 & 10.824 & 5.747  & 1.936  & Transition Disk \\
135 & 13.1981 & 0.0108  & J181414.38-172858.1 & 11.703 & 11.179 & 6.253  & 1.919  & Transition Disk \\
136 & 13.2055 & 0.0025  & J181417.10-172849.0 & 11.824 & 11.332 & 7.058  & 3.638  & Transition Disk \\
137 & 13.2126 & 0.0194  & J181414.22-172757.5 & 12.147 & 11.623 & 6.538  & 1.997  & Transition Disk \\
138 & 13.1858 & 0.0049  & J181414.20-172947.0 & 10.778 & 10.507 & 7.445  & 4.071  & Transition Disk \\
139 & 13.1303 & 0.0691  & J181353.35-173052.1 & 9.467  & 9.299  & 7.496  & 4.097  & Transition Disk \\
140 & 13.1286 & 0.0610  & J181354.93-173111.4 & 10.033 & 9.571  & 4.676  & 2.042  & Transition Disk \\
141 & 13.1316 & 0.0521  & J181357.26-173117.1 & 10.701 & 10.087 & 6.685  & 3.236  & Transition Disk \\
142 & 13.1122 & 0.0773  & J181349.34-173135.1 & 10.111 & 9.883  & 10.262 & 7.892  & Transition Disk \\
143 & 13.1187 & 0.0564  & J181354.75-173150.5 & 11.499 & 11.298 & 6.807  & 4.052  & Transition Disk \\
144 & 13.1589 & 0.0387  & J181403.50-173014.0 & 9.631  & 8.931  & 5.122  & 0.934  & Transition Disk \\
145 & 13.1421 & 0.0526  & J181358.42-173043.0 & 11.134 & 10.699 & 6.780  & 3.402  & Transition Disk \\
146 & 13.1280 & 0.0967  & J181346.96-173011.5 & 9.944  & 9.647  & 8.528  & 6.409  & Transition Disk \\
147 & 13.1250 & 0.0795  & J181350.40-173050.9 & 8.393  & 8.130  & 7.594  & 5.493  & Transition Disk \\
148 & 13.1524 & 0.0761  & J181354.44-172930.1 & 8.655  & 8.281  & 7.398  & 5.147  & Transition Disk \\
149 & 13.1753 & 0.0488  & J181403.24-172904.5 & 9.535  & 8.937  & 2.135  & -2.611 & Transition Disk \\
150 & 13.1832 & 0.0570  & J181402.39-172825.6 & 10.112 & 9.869  & 3.703  & -0.608 & Transition Disk \\
151 & 13.1712 & 0.0418  & J181404.30-172929.5 & 10.602 & 10.088 & 3.017  & -1.599 & Transition Disk \\
152 & 13.1826 & 0.0528  & J181403.24-172834.6 & 9.907  & 9.260  & 1.655  & -2.432 & Transition Disk \\
153 & 13.1639 & 0.0376  & J181404.35-173000.0 & 10.127 & 9.568  & 3.913  & -0.276 & Transition Disk \\
154 & 13.1713 & 0.0395  & J181404.81-172933.3 & 11.091 & 10.374 & 3.963  & -0.653 & Transition Disk \\
155 & 13.2105 & 0.0229  & J181413.21-172758.0 & 11.888 & 11.416 & 5.427  & 1.524  & Transition Disk \\
156 & 13.2038 & 0.0461  & J181407.28-172739.2 & 10.341 & 9.920  & 4.262  & -0.182 & Transition Disk \\
157 & 13.2089 & 0.0266  & J181412.20-172756.8 & 11.358 & 10.652 & 5.704  & 0.822  & Transition Disk \\
158 & 13.2130 & 0.0355  & J181410.72-172728.4 & 10.631 & 10.347 & 4.588  & 1.039  & Transition Disk \\
159 & 13.2103 & 0.0242  & J181412.89-172756.1 & 11.682 & 10.981 & 5.123  & 1.075  & Transition Disk \\
160 & 13.1939 & 0.0384  & J181407.78-172823.8 & 10.681 & 9.957  & 2.256  & -2.989 & Transition Disk \\
161 & 13.1896 & 0.0791  & J181358.26-172727.3 & 10.666 & 10.425 & 6.035  & 3.807  & Transition Disk \\
162 & 13.1801 & 0.0761  & J181357.79-172802.5 & 9.975  & 9.662  & 6.309  & 3.485  & Transition Disk \\
163 & 13.1976 & 0.0803  & J181358.98-172659.8 & 11.393 & 10.948 & 7.773  & 3.752  & Transition Disk \\
164 & 13.1915 & 0.0868  & J181356.80-172708.0 & 11.358 & 10.833 & 6.953  & 4.164  & Transition Disk \\
165 & 13.1984 & 0.0722  & J181400.85-172711.2 & 11.740 & 11.317 & 6.627  & 3.467  & Transition Disk \\
166 & 13.1974 & 0.0662  & J181402.06-172724.8 & 10.170 & 9.973  & 5.734  & 1.855  & Transition Disk \\
167 & 13.2222 & 0.0750  & J181403.09-172551.2 & 9.327  & 8.771  & 8.291  & 2.102  & Transition Disk \\
168 & 13.2081 & 0.0561  & J181405.57-172708.5 & 10.739 & 10.058 & 4.434  & 0.552  & Transition Disk \\
169 & 13.2017 & -0.0300 & J181423.83-172956.8 & 11.641 & 11.337 & 7.013  & 4.566  & Transition Disk \\
170 & 13.2052 & -0.0235 & J181422.82-172934.6 & 11.378 & 11.212 & 7.666  & 5.331  & Transition Disk \\
171 & 13.2266 & -0.0156 & J181423.66-172813.2 & 11.528 & 11.375 & 8.565  & 5.659  & Transition Disk \\
172 & 13.2147 & -0.0087 & J181420.69-172839.2 & 11.950 & 11.775 & 8.072  & 5.042  & Transition Disk \\
173 & 13.2250 & 0.0111  & J181417.54-172732.5 & 10.599 & 10.406 & 8.002  & 5.212  & Transition Disk \\
174 & 13.2556 & -0.0021 & J181424.15-172618.2 & 10.899 & 10.461 & 9.358  & 5.608  &	Transition Disk \\
175 & 13.2439 & 0.0016  & J181421.94-172648.9 & 10.058 & 9.892  & 6.862  & 4.494  & Transition Disk \\
176 & 13.2406 & 0.0040  & J181421.00-172655.2 & 11.241 & 11.043 & 6.649  & 4.894  & Transition Disk \\
177 & 13.2484 & 0.0094  & J181420.74-172621.2 & 11.502 & 11.349 & 7.695  & 5.089  & Transition Disk \\
178 & 13.2676 & 0.0133  & J181422.20-172514.1 & 9.564  & 9.109  & 8.739  & 6.273  & Transition Disk \\
179 & 13.2318 & 0.0604  & J181407.47-172546.1 & 9.299  & 9.099  & 7.004  & 4.299  & Transition Disk \\
180 & 13.2297 & 0.0737  & J181404.30-172529.9 & 9.594  & 9.317  & 4.814  & 1.510  & Transition Disk \\
181 & 13.2416 & 0.0659  & J181407.44-172505.7 & 10.003 & 9.658  & 6.606  & 3.980  & Transition Disk \\
182 & 13.2338 & 0.0719  & J181405.18-172519.8 & 12.536 & 11.969 & 5.976  & 1.548  & Transition Disk \\
183 & 13.2315 & 0.0551  & J181408.63-172556.0 & 10.239 & 10.048 & 7.724  & 5.911  & Transition Disk \\
184 & 13.2344 & 0.0724  & J181405.15-172517.3 & 12.369 & 11.834 & 5.563  & 1.525  & Transition Disk \\
185 & 13.2363 & 0.0654  & J181406.92-172523.1 & 10.862 & 10.665 & 6.555  & 3.011  & Transition Disk \\
186 & 13.1462 & 0.1193  & J181344.17-172835.2 & 9.092  & 8.479  & 7.845  & 5.891  & Transition Disk \\
187 & 13.1188 & 0.1113  & J181342.64-173015.7 & 10.204 & 9.787  & 8.924  & 6.601  & Transition Disk \\
188 & 13.1502 & 0.1042  & J181347.97-172848.4 & 9.896  & 9.186  & 8.720  & 6.425  & Transition Disk \\
189 & 13.1501 & 0.1144  & J181345.71-172831.4 & 9.305  & 9.117  & 8.257  & 6.510  & Transition Disk \\
190 & 13.1924 & 0.0975  & J181354.55-172646.7 & 9.473  & 9.250  & 8.793  & 4.548  & Transition Disk \\
191 & 13.1979 & 0.1120  & J181352.00-172604.4 & 8.698  & 8.412  & 8.768  & 6.937  & Transition Disk \\
192 & 13.1957 & 0.0886  & J181356.90-172651.6 & 11.635 & 11.142 & 6.151  & 3.423  & Transition Disk \\
193 & 13.2292 & 0.0910  & J181400.42-172501.5 & 11.600 & 11.348 & 6.060  & 1.161  & Transition Disk \\
194 & 13.2239 & 0.0903  & J181359.92-172519.4 & 12.463 & 11.933 & 7.178  & 1.502  & Transition Disk \\
195 & 13.2226 & 0.1037  & J181356.81-172500.8 & 11.410 & 10.976 & 6.355  & 3.515  & Transition Disk \\
196 & 13.2203 & 0.1024  & J181356.83-172510.0 & 10.970 & 10.501 & 5.847  & 3.271  & Transition Disk \\
197 & 13.2295 & 0.0835  & J181402.10-172513.4 & 11.661 & 11.143 & 5.267  & 0.385  & Transition Disk \\
198 & 13.2246 & 0.1026  & J181357.29-172456.0 & 10.644 & 10.153 & 5.588  & 3.409  & Transition Disk \\
199 & 13.2298 & 0.0885  & J181401.03-172504.1 & 11.742 & 11.114 & 5.584  & 1.075  & Transition Disk \\
200 & 13.2252 & 0.0813  & J181402.07-172530.9 & 10.578 & 10.153 & 6.911  & 0.759  & Transition Disk \\
201 & 13.2284 & 0.1016  & J181357.97-172445.8 & 12.267 & 11.619 & 5.539  & 3.212  & Transition Disk \\
202 & 13.2280 & 0.0838  & J181401.85-172517.7 & 11.410 & 11.135 & 5.899  & 0.356  & Transition Disk \\
203 & 13.2127 & 0.0831  & J181400.16-172607.3 & 8.735  & 8.405  & 7.955  & 4.510  & Transition Disk \\
204 & 13.2229 & 0.0911  & J181359.62-172521.4 & 11.294 & 10.889 & 7.497  & 2.031  & Transition Disk \\
205 & 13.2372 & 0.1141  & J181356.27-172356.5 & 8.749  & 8.517  & 7.132  & 4.033  & Transition Disk \\
206 & 13.2136 & 0.1166  & J181352.87-172507.0 & 11.056 & 10.724 & 7.227  & 4.170  & Transition Disk \\
207 & 13.2205 & 0.1102  & J181355.12-172455.8 & 10.310 & 10.150 & 7.548  & 4.420  & Transition Disk \\
208 & 13.2377 & 0.0983  & J181359.81-172422.0 & 12.059 & 11.550 & 6.093  & 3.520  & Transition Disk \\
209 & 13.2404 & 0.0764  & J181404.99-172451.4 & 9.730  & 9.387  & 4.296  & 1.688  & Transition Disk \\
210 & 13.2385 & 0.0818  & J181403.55-172448.1 & 10.547 & 9.932  & 4.625  & 1.282  & Transition Disk \\
211 & 13.2315 & 0.0864  & J181401.70-172502.2 & 12.504 & 12.291 & 5.677  & 0.618  & Transition Disk \\
212 & 13.2323 & 0.0929  & J181400.37-172448.5 & 11.057 & 10.655 & 6.449  & 2.195  & Transition Disk \\
213 & 13.2312 & 0.0824  & J181402.55-172510.2 & 10.613 & 9.946  & 4.475  & -0.268 & Transition Disk \\
214 & 13.2292 & 0.0787  & J181403.12-172522.9 & 9.237  & 9.014  & 3.736  & 0.416  & Transition Disk \\
215 & 13.2371 & 0.0936  & J181400.78-172432.1 & 10.320 & 9.876  & 5.402  & 2.454  & Transition Disk \\
216 & 13.2325 & 0.0770  & J181403.89-172515.4 & 10.064 & 9.692  & 5.063  & -0.432 & Transition Disk \\
217 & 13.2404 & 0.0940  & J181401.08-172421.0 & 9.943  & 9.723  & 4.805  & 3.131  & Transition Disk \\
218 & 13.2435 & 0.0771  & J181405.21-172440.2 & 11.152 & 10.697 & 6.127  & 2.319  & Transition Disk \\
219 & 13.2468 & 0.0740  & J181406.27-172435.2 & 11.786 & 11.424 & 7.352  & 4.218  & Transition Disk \\
220 & 13.2352 & 0.0992  & J181359.33-172428.5 & 9.378  & 9.056  & 5.712  & 2.634  & Transition Disk \\
221 & 13.2516 & 0.0858  & J181404.24-172359.7 & 8.614  & 8.314  & 7.176  & 4.687  & Transition Disk \\
222 & 13.2427 & 0.0814  & J181404.14-172435.4 & 10.319 & 10.027 & 4.899  & 2.521  & Transition Disk \\
223 & 13.2390 & 0.0928  & J181401.19-172427.4 & 9.106  & 8.818  & 5.859  & 2.362  & Transition Disk \\
224 & 13.2471 & 0.0781  & J181405.40-172427.1 & 10.256 & 10.098 & 6.963  & 3.952  & Transition Disk \\
225 & 13.2354 & 0.0725  & J181405.22-172513.9 & 10.123 & 9.731  & 4.448  & 1.682  & Transition Disk \\
226 & 13.2422 & 0.0887  & J181402.49-172424.4 & 10.677 & 10.362 & 5.643  & 2.381  & Transition Disk \\
227 & 13.2676 & 0.0821  & J181407.00-172315.7 & 10.238 & 10.018 & 7.799  & 5.071  & Transition Disk \\
228 & 13.2696 & 0.0742  & J181408.98-172322.8 & 10.599 & 10.386 & 7.460  & 4.081  & Transition Disk \\
229 & 13.2668 & 0.0850  & J181406.25-172313.1 & 10.240 & 9.574  & 8.588  & 5.451  & Transition Disk \\
230 & 13.2589 & 0.0796  & J181406.49-172347.4 & 7.302  & 6.825  & 6.675  & 4.609  & Transition Disk \\
231 & 13.2457 & 0.1019  & J181400.00-172350.7 & 9.816  & 9.327  & 10.194 & 5.108  & Transition Disk \\
232 & 13.2399 & 0.1093  & J181357.66-172356.3 & 9.114  & 8.881  & 7.718  & 4.412  & Transition Disk \\
233 & 13.2487 & 0.1005  & J181400.65-172343.4 & 10.996 & 10.360 & 11.104 & 6.363  & Transition Disk \\
234 & 13.2372 & 0.1034  & J181358.63-172415.1 & 9.272  & 8.958  & 6.258  & 3.162  & Transition Disk \\

\enddata

\end{deluxetable}

\clearpage

% Fig 12

% Fig 13

% Fig 14
\begin{figure*}[ht]
\begin{center}
% \setcaptionmargin{1cm}
\includegraphics[scale=0.47,angle=0]{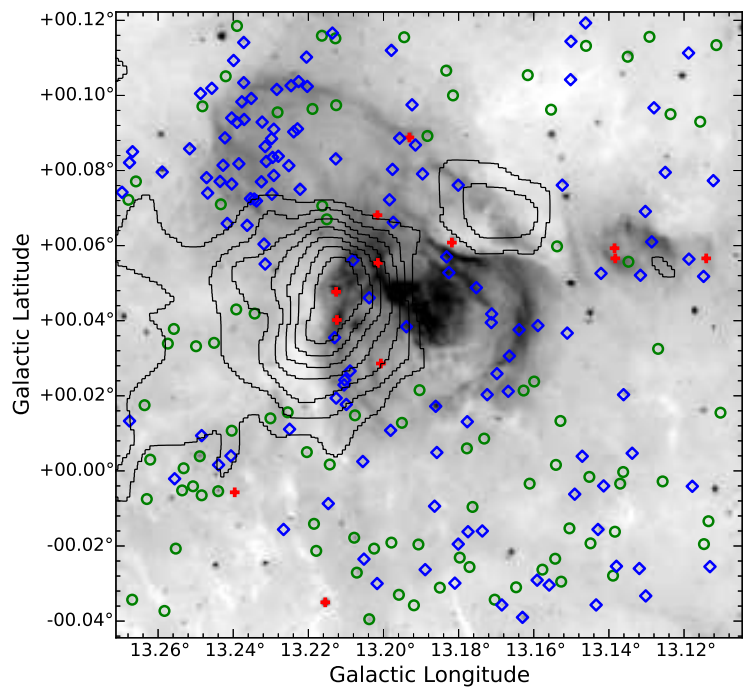}
\caption {Distribution of the identified YSOs. Background is Spitzer 8.0 $\mu$m image. Red crosses are Class I sources; green circles are class II sources; blue diamonds are "transition disk" sources. Black contours shows the $^{13}$CO emission around the bubble.} 
\label{yso_distribution}
\end{center}
\end{figure*}

\clearpage

%__________________________________________________________________

\section{Discussion}
\label{discussion}

%________________________________________________________________

In this section we discuss the distribution of molecular material around the bubble N10 and its connection with the star formation history. 

\subsection{Surrounding gas of N10}
\label{surrounding-gas}

Channel maps are presented in Figure \ref{channel_map}; since $^{12}$CO is optically thick, it appears to be spread over a large area, whereas $^{13}$CO traces the denser regions, once it is optically thinner. Figure \ref{clumps} displays two peaks of molecular emission, i.e. two $^{13}$CO condensations. Clump \#1 is centered at $l$ = 13.218$^{\circ}$, $b$ = 0.043$^{\circ}$ and Clump \#2 is located at $l$ = 13.169$^{\circ}$, $b$ = 0.072$^{\circ}$. The two clumps are located precisely on the edge of the ring structure revealed by the 8.0 $\mu$m emission and highlighted by an ellipse in Figure \ref{map8_objects}. 

In order to verify the dynamical status of the clumps, we compare the gas mass in LTE and the virial mass calculated in Subsection \ref{obs-deriv-param} for each clump. The two clumps have greater M$_{virial}$ than M$_{LTE}$, implying that they are gravitationally unbound, indicating currently there is no star in forming \citep[e.g.][]{jinghua2014,liu2016}.

\subsection{Other components of CO emission}
\label{Other components}

As highlighted in subsection \ref{Molecular Emission} CO components at 20 and 37 km s$^{-1}$ do not seem to be physically related with bubble \object{N10}. Which could be the origin of this contribution? It is known CO emission in galaxies is concentrated in spiral arms \citep{nieten2006,schinnerer2013}. The line of sight towards \object{N10} crosses two spiral arms before reaching the distance 4.7 kpc \citep[see e.g. Figure 10 of][]{hou2014}. The velocities of peaks 20 and 37 km s$^{-1}$ correspond to near kinematic distances of 2.4 and 3.7 kpc respectively, representing roughly to the distance of those arms. 

It is therefore reasonable to suppose these two velocity peaks are associated with foreground gas situated in distinct spiral arms. Furthermore, emission at these two velocities does not seem to be correlated with geometry of \object{N10}; emission appears to be irregularly spread over studied field. If one examines the Figure 4 of the $^{13}$CO (1--0) survey of the Galaxy by \cite{lee2001}, selecting the panel corresponding to $b$ = 0.05$^{\circ}$, one can see at longitude 13.2$^{\circ}$ the presence of $^{13}$CO at about 50, 35 and 20 km s$^{-1}$. The last two ones are part of elongated structures in the longitude-velocity diagram (extending to higher and lower longitudes) that are usually interpreted as spiral arms. In this region of the diagram, lower velocities correspond to closest arms.

Note that kinematic distances are uncertain at longitudes close to the Galactic center. We have to make use of a $^{13}$CO survey because in $^{12}$CO longitude-velocity diagrams, the spiral arms are wider in velocity, and are not seen separated.

\subsection{Situation of star formation}
\label{situation-sf}

The densest clump in the 870 $\mu$m emission seems to be a candidate region to form stellar clusters, since it have a total mass of $M_{tot} = 240$ M$_{\odot}$ and a mean radius of $R = 0.36$ pc. In accordance with \cite{motte2003} fragments in the range between 0.09 and 0.56 pc and masses covering a range from 20 to 3600 M$_{\odot}$ have characteristics of protoclusters.

\cite{elmegreen1977} was the first to propose the scenario of ``Collect and Collapse'' where the radiation of the massive stars of an HII region creates an ionization front at the interface with the molecular cloud, that drives the propagation of a shock front into the neutral material and which accumulates mass and eventually becomes gravitationally unstable. Other scenarios of triggered star formation have been proposed, like e.g. the ``Radiation-Driven Implosion'' model, based on the over pressure exerted by the ionized gas, suggested by \cite{lefloch1994}. While ``Collect and Collapse'' model takes place in a large spatial size ($\sim$10 pc) with a longer timescale (a few Myr),``Radiation-Driven Implosion'' takes place in $\sim$ 1 pc with a timescale of 0.5 Myr. 

Although we found evidences for active star formation in N10, we are not sure that the formation of these YSOs were triggered by the ``Collect and Collapse'' mechanism around the infrared bubble. In order to verify if this process is viable we can apply the analytical model proposed by \cite{whitworth1994} and compare the fragmentation time scale $t_{frag}$ with the dynamical age $t_{dyn}$ of the region. The \cite{whitworth1994} model describes the fragmentation time as:

\begin{eqnarray} \label{t-frag}
\mathrm t_{frag} = 1.56 \ c_{s}^{7/11} \ N_{uv} ^{-1/11} n_{o}^{-5/11},
\end{eqnarray}

\noindent where $c_{s}$ is the isothermal sound speed in the ionized gas in the shocked layer in units of $0.2$ km s$^{-1}$, $N_{uv}$ is the ionizing photon flux in units of 10$^{49}$ photons.s$^{-1}$ and $n_{o}$ is the initial particle number density of the ambient neutral gas in units of 10$^{3}$ cm$^{-3}$. Considering $c_{s} = 0.2$ km s$^{-1}$ \citep{liu2012}, $N_{uv} = 1.86 \times 10^{49}$ photons.s$^{-1}$ and $n_{o} \sim 10^{3}$ cm$^{-3}$ \citep{ma2013} we estimated $t_{frag} \sim 1.5 \times 10^{6}$ yr for the region. From \cite{ma2013} $t_{dyn} = 9.17 \times 10^{4}$ yr, i.e. the dynamical age is smaller than the fragmentation time scale, which indicates that the region do not support the ``Collect and Collapse'' mechanism. In this case the ``Radiation-Driven Implosion'' could be considered and further investigated.

The position of YSOs compared with CO distribution indicates that stars are forming inside the molecular clumps. Figure \ref{yso_distribution} shows the spatial distribution of identified YSOs from the Table \ref{all_ysos}. In fact the Class I YSOs candidates to be associated to N10 presented in Table \ref{model-parameters} have ages smaller than the fragmentation time scale, suggesting a possibility of triggered star formation by pre-existing condensations compressed by the pressure of the ionized gas, as ``Radiation-Driven Implosion'' scenario proposes.

\subsection{The bubble N11}

Infrared images shows N11, a bubble that seems to be physically connected to \object{N10} and extends about 3 pc in the up-right direction. Conversely the molecular distribution of N11 does not suggest a physical connection with \object{N10}, since the emission of $^{13}$CO (1--0) between 47 and 53 km s$^{-1}$ is not coincident with 8.0 $\mu$m emission. 

It is probable that this object is a remnant of an HII region, where the lack of 20 cm emission lead us tho consider there is no more ionized gas inside. It is likely that other energy source has triggered the formation of these YSOs, such as the explosion of a type II supernova.

Class I YSOs do not seem to be superimposed on the bubble N11 and many Class II YSOs can be found towards N11, as we can see in the Figure \ref{yso_distribution}. There is a remarkable concentration of transition disk sources surrounding the top frontier of N11. We consider, in this interpretation, that the concentration of Class II YSOs near the upper frontier of N11 is possibly the result of a past star formation activity related to that bubble.

\subsection{A small bubble to the right of N10: \\ MWP1G013134+000580}
\label{A small bubble}

The bubble MWP1G013134+000580, at coordinates $l$ = 13.134$^{\circ}$ and $b$ = 0.058$^{\circ}$, has size smaller than 2 pc. Interestingly this small bubble should have the about the same distance of \object{N10}, since its CO emission is contained in the same main velocity peak, clearly seen in channel maps with velocities between 51 and 53 km s$^{-1}$ in Figure \ref{channel_map}.

We found three Class I YSOs in the region covered by 8.0 $\mu$m emission of MWP1G013134+000580. The age of the small bubble seems to have the same order of \object{N10}, as we can infer from the evolutionary stages of the YSOs.

%__________________________________________________________________

\section{Conclusions}
\label{conclusions}

%__________________________________________________________________

We have performed a comprehensive study of the infrared bubble \object{N10} using the molecular line emissions of $^{12}$CO ($J=1-0$) and $^{13}$CO ($J=1-0$), mid-infrared Spitzer-GLIMPSE and MIPSGAL images, VLA data of the 20 cm emission, APEX observations of the continuum 870 $\mu$m emission and WISE catalog of mid-infrared point sources. The key results are summarized as follows:

\begin{enumerate}
	\item We observed the $J = 1 - 0$ transition of CO isotopologes at PMO 13.7-m radio telescope. The distribution of the CO emission showed that the molecular gas around the bubble N10 has velocity $V_{lsr} = 52.6$ km s$^{-1}$, from which we estimated distance $D = 4.7 \pm 0.5$ kpc. This observations revealed two $^{13}$CO clumps with $M_{LTE} \sim 2 \times 10^{3}$ M$_{\odot}$, $M_{virial} \sim 8.5 \times 10^{3}$ M$_{\odot}$ and $M_{Jeans} \sim 6 \times 10^{3}$ M$_{\odot}$, which means that the clumps implying that they are gravitationally unbound currently.

\item The emission of radio continuum and the presence of 24 $\mu$m emission suggest ionizing sources inside the bubble. We estimated a total flux of 20 cm of $F_{20\ cm} = 1.17$ Jy and an electron density of $n_{e} \sim 130$ cm$^{-3}$, with a Lyman continuum photon flux of $N_{uv} = 1.86 \times 10^{49}$ ionizing photons s$^{-1}$, equivalent to an $\sim$ O7 V star (or stars) keeping the gas ionized.
    
    \item Two cold dust clumps were identified towards N10 in LABOCA/APEX images. For the densest clump, we estimated from emission at 870 $\mu$m a total mass of $M_{tot} = 240$ M$_{\odot}$, a mean radius of $R_{D} = 0.36$ pc, a column density of $N(H_{2}) = 6.3 \times 10^{22}$ cm$^{-2}$ and an average volume density of $n(H_{2}) = 9.4 \times \ 10^{4}$ cm $^{-3}$, physical characteristics indicating that this condensation is a good candidate of protocluster.
    
    \item We identified 234 YSOs in the whole region: 12 of them classified as Class I, 91 Class II and 131 Transition Disks. We fitted the SED for Class I YSOs candidates identified from \#1 to \#9 and we derived their physical parameters. From the models we found stellar ages ranging from $\sim 10^{3}$ to $10^{6}$ yr. By comparing the estimated dynamical age ($t_{dyn} = 9.17 \times 10^{4}$ yr) and the fragmentation time scale ($t_{frag} \sim 1.5 \times 10^{6}$ yr) we infer that star formation can be triggered as a consequence of the ``Radiation-Driven Implosion'' process. Likewise, the age range for the Class I YSOs are below that found for the fragmentation time scale, indicating they were formed before the collect molecular cloud became gravitationally unstable to fragment to form stars.
    
    \item In the Spitzer 8.0 $\mu$m image the infrared bubble N11 can be seen in the direction of the N10, however one is not physically connected with other. Class II YSOs appears towards N11, suggesting that this could be a remnant of HII region. A third infrared bubble, the small MWP1G013134+000580, appears in the observed field and, interestingly, has CO emission in the same main velocity as N10 and seems to shelter some evolved YSOs.
    
\end{enumerate}

\acknowledgments

      We gratefully acknowledge the contribution of Tie Liu, who provided some of the scripts used in this work. We also give our thanks Cristina Cappa and Bertrand Lefloch for the helpful discussion. We thank the anonymous referee for useful comments and suggestions that led to a improved version of the original article. This work is supported by the National Natural Science Foundation of China through grant NSFC 11373009-11433008. J.Y is supported by the National Natural Science Foundation of China through grants of 11503035 and 11573036. EM acknowledges support from the Brazilian agency FAPESP under the grants 2014/22095-6 and 2015/22254-0. We are grateful to the staffs at the Qinghai Station of PMO for their hospitality and assistance during the observations. We thank the Key Laboratory for Radio Astronomy, CAS, for partial support in the operation of the telescope.


\begin{thebibliography}{}

\bibitem[Am{\^o}res \& L{\'e}pine(2005)]{amores2005} Am{\^o}res, E.~B., \& L{\'e}pine, J.~R.~D.\ 2005, \aj, 130, 659 

\bibitem[Anderson et al.(2014)]{anderson2014} Anderson, L.~D., Bania, T.~M., Balser, D.~S., et al.\ 2014, \apjs, 212, 1 

\bibitem[Beaumont \& Williams(2010)]{beaumont2010} Beaumont, C.~N., \& Williams, J.~P.\ 2010, \apj, 709, 791 

\bibitem[Becker et al.(1994)]{becker1994} Becker, R.~H., White, R.~L., Helfand, D.~J., \& Zoonematkermani, S.\ 1994, \apjs, 91, 347 

\bibitem[Benjamin et al.(2003)]{benjamin2003} Benjamin, R.~A., Churchwell, E., Babler, B.~L., et al.\ 2003, \pasp, 115, 953 

\bibitem[Bergin \& Tafalla(2007)]{bergin2007} Bergin, E.~A., \& Tafalla, M.\ 2007, \araa, 45, 339 

\bibitem[Beuther et al.(2011)]{beuther2011} Beuther, H., Linz, H., Henning, T., et al.\ 2011, \aap, 531, AA26 

\bibitem[Blaauw(1991)]{blaauw1991} Blaauw, A.\ 1991, NATO ASIC Proc.~342: The Physics of Star Formation and Early Stellar Evolution, 125 

\bibitem[Brand \& Blitz(1993)]{brand1993} Brand, J., \& Blitz, L.\ 1993, \aap, 275, 67 

\bibitem[Bressert et al.(2010)]{bressert2010} Bressert, E., Bastian, N., Gutermuth, R., et al.\ 2010, \mnras, 409, L54 

\bibitem[Cappa et al.(2009)]{cappa2009} Cappa, C.~E., Rubio, M., Mart{\'{\i}}n, M.~C., \& Romero, G.~A.\ 2009, \aap, 508, 759 

% \bibitem[Cappa et al.(2014)]{cappa2014} Cappa, C.~E., Rubio, M., Romero, G.~A., Duronea, N.~U., \& Firpo, V.\ 2014, \aap, 562, A6 

\bibitem[Cappa et al.(2016)]{cappa2016} Cappa, C.~E., Duronea, N., Firpo, V., et al.\ 2016, \aap, 585, A30 

\bibitem[Churchwell et al.(2006)]{churchwell2006} Churchwell, E., Povich, M.~S., Allen, D., et al.\ 2006, \apj, 649, 759 

\bibitem[Churchwell et al.(2007)]{churchwell2007} Churchwell, E., Watson, D.~F., Povich, M.~S., et al.\ 2007, \apj, 670, 428 

\bibitem[Churchwell et al.(2009)]{churchwell2009} Churchwell, E., Babler, B.~L., Meade, M.~R., et al.\ 2009, \pasp, 121, 213 

\bibitem[Condon et al.(1998)]{condon1998} Condon, J.~J., Cotton, W.~D., Greisen, E.~W., et al.\ 1998, \aj, 115, 1693 

\bibitem[Cutri \& IPAC/WISE Science Data Center Team(2011)]{cutri2011} Cutri, R.~M., \& IPAC/WISE Science Data Center Team 2011, Bulletin of the American Astronomical Society, 43, \#301.02 

\bibitem[Cutri et al.(2013)]{cutri2013} Cutri, R.~M., et al.\ 2013, VizieR Online Data Catalog, 2328, 0 

\bibitem[Dale et al.(2005)]{dale2005} Dale, J.~E., Bonnell, I.~A., Clarke, C.~J., \& Bate, M.~R.\ 2005, \mnras, 358, 291 

\bibitem[Dale \& Bonnell(2008)]{dale2008} Dale, J.~E., \& Bonnell, I.~A.\ 2008, \mnras, 391, 2 

\bibitem[Deharveng et al.(2008)]{deharveng2008} Deharveng, L., Lefloch, B., Kurtz, S., et al.\ 2008, \aap, 482, 585 

\bibitem[Deharveng et al.(2010)]{deharveng2010} Deharveng, L., Schuller, F., Anderson, L.~D., et al.\ 2010, \aap, 523, A6 

\bibitem[Dewangan et al.(2012)]{dewangan2012} Dewangan, L.~K., Ojha, D.~K., Anandarao, B.~G., Ghosh, S.~K., \& Chakraborti, S.\ 2012, \apj, 756, 151 

\bibitem[Deharveng et al.(2015)]{deharveng2015} Deharveng, L., Zavagno, A., Samal, M.~R., et al.\ 2015, \aap, 582, A1 

\bibitem[Dewangan \& Ojha(2013)]{dewangan2013} Dewangan, L.~K., \& Ojha, D.~K.\ 2013, \mnras, 429, 1386

\bibitem[Dewangan et al.(2015)]{dewangan2015a} Dewangan, L.~K., Ojha, D.~K., Grave, J.~M.~C., \& Mallick, K.~K.\ 2015, \mnras, 446, 2640 

\bibitem[Draine \& Anderson(1985)]{draine1985} Draine, B.~T., \& Anderson, N.\ 1985, \apj, 292, 494 

\bibitem[Duronea et al.(2015)]{duronea2015} Duronea, N.~U., Vasquez, J., G{\'o}mez, L., et al.\ 2015, \aap, 582, A2 

\bibitem[Dyson \& Williams(1980)]{dyson1980} Dyson, J.~E., \& Williams, D.~A.\ 1980, New York, Halsted Press, 1980.~204 p.,  

\bibitem[Elmegreen \& Lada(1977)]{elmegreen1977} Elmegreen, B.~G., \& Lada, C.~J.\ 1977, \apj, 214, 725 

\bibitem[Fazio et al.(2004)]{fazio2004} Fazio, G.~G., Hora, J.~L., Allen, L.~E., et al.\ 2004, \apjs, 154, 10 

\bibitem[Feigelson \& Montmerle(1999)]{feigelson1999} Feigelson, E.~D., \& Montmerle, T.\ 1999, \araa, 37, 363   

\bibitem[Garden et al.(1991)]{garden1991} Garden, R.~P., Hayashi, M., Hasegawa, T., Gatley, I., \& Kaifu, N.\ 1991, \apj, 374, 540 

\bibitem[Gutermuth et al.(2009)]{gutermuth2009} Gutermuth, R.~A., Megeath, S.~T., Myers, P.~C., et al.\ 2009, \apjs, 184, 18

\bibitem[Helfand et al.(2006)]{helfand2006} Helfand, D.~J., Becker, R.~H., White, R.~L., Fallon, A., \& Tuttle, S.\ 2006, \aj, 131, 2525 

\bibitem[Hildebrand(1983)]{hildebrand1983} Hildebrand, R.~H.\ 1983, \qjras, 24, 267 

\bibitem[Hou \& Han(2014)]{hou2014} Hou, L.~G., \& Han, J.~L.\ 2014, \aap, 569, AA125 

\bibitem[Huang(1954)]{huang1954} Huang, S.-S.\ 1954, \aj, 59, 137 

\bibitem[Ji et al.(2012)]{ji2012} Ji, W.-G., Zhou, J.-J., Esimbek, J., et al.\ 2012, \aap, 544, A39 

\bibitem[Kendrew et al.(2012)]{kendrew2012} Kendrew, S., Simpson, R., Bressert, E., et al.\ 2012, \apj, 755, 71 

\bibitem[Kendrew et al.(2016)]{kendrew2016} Kendrew, S., Zieleniewski, S., Houghton, R.~C.~W., et al.\ 2016, \mnras, 458, 2405 

\bibitem[Koenig et al.(2012)]{koenig2012} Koenig, X.~P., Leisawitz, D.~T., Benford, D.~J., et al.\ 2012, \apj, 744, 130  

\bibitem[Koenig \& Leisawitz(2014)]{koenigleisawitz2014} Koenig, X.~P., \& Leisawitz, D.~T.\ 2014, \apj, 791, 131 

\bibitem[Lee et al.(2001)]{lee2001} Lee, J.-K., Walsh, A.~J., Burton, M.~G., \& Ashley, M.~C.~B.\ 2001, \mnras, 324, 1102 

\bibitem[Lefloch \& Lazareff(1994)]{lefloch1994} Lefloch, B., \& Lazareff, B.\ 1994, \aap, 289, 559 

\bibitem[Lefloch et al.(2002)]{lefloch2002} Lefloch, B., Cernicharo, J., Rodr{\'{\i}}guez, L.~F., et al.\ 2002, \apj, 581, 335 

\bibitem[Lefloch et al.(2005)]{lefloch2005} Lefloch, B., Cernicharo, J., Cabrit, S., \& Cesarsky, D.\ 2005, \aap, 433, 217 

\bibitem[Liu et al.(2012)]{liu2012} Liu, T., Wu, Y., Zhang, H., \& Qin, S.-L.\ 2012, \apj, 751, 68 

\bibitem[Liu et al.(2015)]{liu2015} Liu, H.-L., Wu, Y., Li, J., et al.\ 2015, \apj, 798, 30 

\bibitem[Liu et al.(2016)]{liu2016} Liu, H.-L., Li, J.-Z., Wu, Y., et al.\ 2016, \apj, 818, 95 

\bibitem[Lockman(1989)]{lockman1989} Lockman, F.~J.\ 1989, \apjs, 71, 469

\bibitem[Ma et al.(2013)]{ma2013} Ma, Y., Zhou, J., Esimbek, J., et al.\ 2013, \apss, 345, 297 

\bibitem[Mathis et al.(1977)]{mathis1977} Mathis, J.~S., Rumpl, W., \& Nordsieck, K.~H.\ 1977, \apj, 217, 425 

\bibitem[Matsakis et al.(1976)]{matsakis1976} Matsakis, D.~N., Evans, N.~J., II, Sato, T., \& Zuckerman, B.\ 1976, \aj, 81, 172 

\bibitem[Minier \& Booth(2002)]{minier2002} Minier, V., \& Booth, R.~S.\ 2002, \aap, 387, 179 

\bibitem[Miettinen(2012)]{miettinen2012} Miettinen, O.\ 2012, \aap, 542, A101 

\bibitem[Motte et al.(2003)]{motte2003} Motte, F., Schilke, P., \& Lis, D.~C.\ 2003, \apj, 582, 277 

\bibitem[Nieten et al.(2006)]{nieten2006} Nieten, C., Neininger, N., Gu{\'e}lin, M., et al.\ 2006, \aap, 453, 459 

\bibitem[Panagia \& Walmsley(1978)]{panagia1978} Panagia, N., \& Walmsley, C.~M.\ 1978, \aap, 70, 411 

\bibitem[Pandian \& Goldsmith(2007)]{pandian2007} Pandian, J.~D., \& Goldsmith, P.~F.\ 2007, \apj, 669, 435 

\bibitem[Pandian et al.(2008)]{pandian2008} Pandian, J.~D., Momjian, E., \& Goldsmith, P.~F.\ 2008, \aap, 486, 191

\bibitem[Paladini et al.(2012)]{paladini2012} Paladini, R., Umana, G., Veneziani, M., et al.\ 2012, \apj, 760, 149 

\bibitem[Pillai et al.(2007)]{pillai2007} Pillai, T., Wyrowski, F., Hatchell, J., Gibb, A.~G., \& Thompson, M.~A.\ 2007, \aap, 467, 207 

\bibitem[Reich et al.(1990)]{reich1990} Reich, W., Fuerst, E., Reich, P., \& Reif, K.\ 1990, \aaps, 85, 633 

\bibitem[Rieke et al.(2004)]{rieke2004} Rieke, G.~H., Young, E.~T., Engelbracht, C.~W., et al.\ 2004, \apjs, 154, 25 

\bibitem[Robitaille et al.(2007)]{robitaille2007} Robitaille, T.~P., Whitney, B.~A., Indebetouw, R., \& Wood, K.\ 2007, \apjs, 169, 328 

\bibitem[Roman-Duval et al.(2009)]{roman2009} Roman-Duval, J., Jackson, J.~M., Heyer, M., et al.\ 2009, \apj, 699, 1153 

\bibitem[Samal et al.(2014)]{samal2014} Samal, M.~R., Zavagno, A., Deharveng, L., et al.\ 2014, \aap, 566, A122 

\bibitem[Sault et al.(1995)]{sault1995} Sault, R.~J., Teuben, P.~J., \& Wright, M.~C.~H.\ 1995, Astronomical Data Analysis Software and Systems IV, 77, 433 

\bibitem[Schinnerer et al.(2013)]{schinnerer2013} Schinnerer, E., 
Meidt, S.~E., Pety, J., et al.\ 2013, \apj, 779, 42 

\bibitem[Schuller et al.(2009)]{schuller2009} Schuller, F., Menten, K.~M., Contreras, Y., et al.\ 2009, \aap, 504, 415 

\bibitem[Shan et al.(2012)]{shan2012} Shan, W., Yang, J., Shi, S., et al.\ 2012, Transactions on Terahertz Science and Technology, 2, 593

\bibitem[Simpson et al.(2012)]{simpson2012} Simpson, R.~J., Povich, M.~S., Kendrew, S., et al.\ 2012, \mnras, 424, 2442 

\bibitem[Stahler \& Palla(2005)]{stahler2005} Stahler, S.~W., \& Palla, F.\ 2005, The Formation of Stars, by Steven W.~Stahler, Francesco Palla, pp.~865.~ISBN 3-527-40559-3.~Wiley-VCH , January 2005

\bibitem[Szymczak et al.(2000)]{szymczak2000} Szymczak, M., Hrynek, G., \& Kus, A.~J.\ 2000, \aaps, 143, 269 

\bibitem[Szymczak et al.(2002)]{szymczak2002} Szymczak, M., Kus, A.~J., Hrynek, G., K{\v e}pa, A., \& Pazderski, E.\ 2002, \aap, 392, 277 

\bibitem[Tackenberg et al.(2012)]{tackenberg2012} Tackenberg, J., Beuther, H., Henning, T., et al.\ 2012, \aap, 540, AA113 

\bibitem[Thompson et al.(2006)]{thompson2006} Thompson, M.~A., Hatchell, J., Walsh, A.~J.,   MacDonald, G.~H., \& Millar, T.~J.\ 2006, \aap, 453, 1003 

\bibitem[Thompson et al.(2012)]{thompson2012} Thompson, M.~A., Urquhart, J.~S., Moore, T.~J.~T., \& Morgan, L.~K.\ 2012, \mnras, 421, 408 

\bibitem[Ungerechts et al.(2000)]{ungerechts2000} Ungerechts, H., Umbanhowar, P., \& Thaddeus, P.\ 2000, \apj, 537, 221 

\bibitem[Watson et al.(2008)]{watson2008} Watson, C., Povich, M.~S., Churchwell, E.~B., et al.\ 2008, \apj, 681, 1341 

\bibitem[Whitworth et al.(1994)]{whitworth1994} Whitworth, A.~P., Bhattal, A.~S., Chapman, S.~J., Disney, M.~J., \& Turner, J.~A.\ 1994, \mnras, 268, 291 


\bibitem[Wienen et al.(2012)]{wienen2012} Wienen, M., Wyrowski, F., Schuller, F., et al.\ 2012, \aap, 544, A146 

\bibitem[Wilson \& Rood(1994)]{wilson1994} Wilson, T.~L., \& Rood, R.\ 1994, \araa, 32, 191 

\bibitem[Wright et al.(2010)]{wright2010} Wright, E.~L., Eisenhardt, P.~R.~M., Mainzer, A.~K., et al.\ 2010, \aj, 140, 1868 

\bibitem[Wu et al.(2012)]{wu2012} Wu, Y., Liu, T., Meng, F., 
et al.\ 2012, \apj, 756, 76 

\bibitem[Yuan et al.(2014)]{jinghua2014} Yuan, J.-H., Wu, Y., Li, J.~Z., \& Liu, H.\ 2014, \apj, 797, 40 

\bibitem[Zavagno et al.(2010)]{zavagno2010} Zavagno, A., Anderson, L.~D., Russeil, D., et al.\ 2010, \aap, 518, L101 

\end{thebibliography}
\end{document}